 \documentclass[a4paper,11pt]{article}
\pdfoutput=1 
 
\usepackage{jheppub}

\usepackage{placeins}
\usepackage{mathrsfs}
\usepackage{wasysym}
\usepackage{verbatim}
\usepackage{booktabs}
\usepackage[tight]{subfigure}
\usepackage{color}
\usepackage{graphicx}
\usepackage{siunitx}

\newcommand{\be}{\begin{equation}}
\newcommand{\ee}{\end{equation}}
\newcommand{\bea}{\begin{eqnarray}}
\newcommand{\eea}{\end{eqnarray}}

\def\simgt{\stackrel{>}{{}_\sim}}
\newcommand{\gsim}{\lower.7ex\hbox{$\;\stackrel{\textstyle>}{\sim}\;$}}
\newcommand{\lsim}{\lower.7ex\hbox{$\;\stackrel{\textstyle<}{\sim}\;$}}

\def\mass2{mass${}^2$}

\def\mass2{mass${}^2$}

\def\simgt{\stackrel{>}{{}_\sim}}

\definecolor{nicered}{rgb}{0.7,0.1,0.1}
\definecolor{nicegreen}{rgb}{0.1,0.5,0.1}
\hypersetup{colorlinks,citecolor= nicegreen,linkcolor= nicered}

\title{Naturalness of MSSM dark matter}
\author[a]{Mar\'{\i}a Eugenia Cabrera,}\author[b]{J. Alberto Casas,} \author[c]{Antonio Delgado,} \author[b, d]{Sandra Robles,} \author[e]{and Roberto Ruiz de Austri } 
\affiliation[a]{Instituto de F\'isica, Universidade de S\~{a}o Paulo, S\~{a}o
  Paulo SP 05508-900, Brazil}
\affiliation[b]{Instituto de F\'isica Te\'orica, IFT-UAM/CSIC, Universidad Aut\'onoma de Madrid, \\
        Cantoblanco, 28049 Madrid, Spain }
\affiliation[c]{Department of Physics, University of Notre Dame, Notre Dame, IN 46556, USA}
\affiliation[d]{Departamento de F\'{\i}sica Te\'{o}rica, Universidad Aut\'{o}noma de Madrid,  \\
        Cantoblanco, 28049 Madrid, Spain }  
\affiliation[e]{Instituto de F\'{i}sica Corpuscular, IFIC-UV/CSIC, Valencia, Spain}        
\emailAdd{mcabrera@if.usp.br} \emailAdd{alberto.casas@uam.es}\emailAdd{antonio.delgado@nd.edu} \emailAdd{sandra.robles@uam.es} \emailAdd{rruiz@ific.uv.es} 

\abstract{\small
  There exists a vast literature examining the electroweak (EW) fine-tuning
  problem in supersymmetric scenarios, but little concerned with the dark
  matter (DM) one, which should be combined with the former.  In this paper,
  we study this problem in an, as much as possible, exhaustive and rigorous
  way. We have considered the MSSM framework, assuming that the LSP is the
  lightest neutralino, $\chi_1^0$, and exploring 
  the various possibilities for the mass and  composition of $\chi_1^0$, 
  as well as different 
  mechanisms for annihilation of the DM particles in the early Universe
  (well-tempered neutralinos, funnels and co-annihilation scenarios). We also
  present a discussion about the statistical meaning of the fine-tuning and
  how it should be computed for the DM abundance, and combined with the EW
  fine-tuning. The results are very robust and model-independent and favour
  some scenarios (like the h-funnel when $M_{\chi_1^0}$ is not too close to
  $m_h/2$) with respect to others (such as the pure wino case). These features
  should be taken into account when one explores ``natural SUSY'' scenarios
  and their possible signatures at the LHC and in DM detection experiments.  }

\preprint{IFT-UAM/CSIC-16-025\\[-8mm]
         \begin{flushright} 
          FTUAM-16-9
          \end{flushright} }

\begin{document}

\maketitle
\flushbottom

\section{Introduction}
\label{sec:intro}
In the minimal supersymmetric Standard Model (MSSM), there are several potential sources of fine-tuning. The most notorious one is the electroweak (EW) fine-tuning, which generically requires light gluino, light Higgsinos, (not so) light winos and, in many cases, light stops. This fine-tuning can be reasonably quantified by the ``standard'' measure \cite{Ellis:1986yg,Barbieri:1987fn}:
\be
\label{BG}
\Delta^{\rm (EW)}_i = \frac{d \log v^2}{d \log \theta_i}\ ;\ \ \ \ \ \ \ \Delta^{\rm (EW)}\equiv\max\left\{\Delta^{\rm (EW)}_i\right\} ,
\ee
where $v^2$ is the Higgs vacuum expectation value (VEV) and $\theta_i$ are the
independent (initial) parameters of the model under consideration. Typically
$\Delta^{\rm (EW)}$ is dominated by the gluino-mass parameter and its value is
$\gsim {\cal O}(100)$ \cite{Papucci:2011wy}, corresponding to a fine-tuning at
the level of $\lsim 1\%$. There is a vast literature concerning this EW
fine-tuning of the MSSM. An important fact is that $\tan\beta$ should be
moderately large (say $\tan\beta\simgt 6$) in order to reproduce the
experimental Higgs mass without the need of gigantic stop masses, which would
imply a very severe fine-tuning.

Besides the EW fine-tuning, there is a potential fine-tuning related to the
generation of the right amount of dark matter (DM). In some scenarios of
supersymmetric dark matter, a delicate balance between a-priori-independent
quantities is required, denoting a fine-tuned situation. Here, by contrast, the literature
is much less extensive \cite{Fichet:2012sn, Grothaus:2012js, Cheung:2012qy, Cohen:2013kna} and, furthermore, many important
mechanisms of supersymmetric dark matter have never been considered from this
point of view\footnote{For works studying the effect of DM constraints on the EW fine-tuning, see e.g. \cite{Boehm:2013gst,Barducci:2015ffa}.}. The main goal of this paper is precisely to perform a rigorous
study of the fine-tuning associated with the production of MSSM dark matter in
all the interesting scenarios.  Moreover, we will combine this fine-tuning
with the EW one, to select the MSSM regions that are globally less fine-tuned.

We will focus on the case where the DM particle is a supersymmetric WIMP, namely the lightest state of the neutralino mass matrix,
\bea
\label{M_X}
{\cal M}_{\chi^0} = \left(
\begin{array}{cccc}
M_1 &  0 & - m_Z s_W c_\beta & m_Z s_W s_\beta \\
0 & M_2  & m_Z c_W c_\beta & - m_Z c_W s_\beta \\
 - m_Z s_W c_\beta & m_Z c_W c_\beta & 0 & -\mu \\
 m_Z s_W s_\beta & - m_Z c_W s_\beta & -\mu & 0 
\end{array}
\right),
\eea
which is the ``standard" situation. Of course, the lightest neutralino, $\chi_1^0$, must also be the lightest supersymmetric particle (LSP). In the previous equation, $M_1$ and $M_2$ are the (low-energy) bino and wino soft mass parameters, while $\mu$ is the mass parameter in the superpotential, which gives mass to Higgsinos. As usual, $s_W$ ($c_W$) is the sin (cosine) of the weak angle and $s_\beta$ ($c_\beta$) is the sin (cosine) of the $\beta-$angle, defined by the ratio of the two Higgs VEVs, $\tan\beta= \langle H_u\rangle/\langle H_d\rangle$. Generically, $\chi_1^0$ is a combination of bino, wino and Higgsinos, though is usually dominated by one of these species. 
Certainly, the content of $\chi_1^0$ in each species depends on the particular values of the four parameters that define ${\cal M}_{\chi^0}$, i.e. $\{M_1,M_2,\mu,\tan\beta\}$.

The  lightest neutralino is a perfect candidate for DM, but, to be successful, it must be produced in the early Universe in the right amount to reproduce the present DM relic density \cite{Ade:2015xua}
\be
\label{OmegaDM}
\Omega_{\rm DM}^{\rm (obs)}h^2 = 0.119 \pm 0.012 \ .
\ee

We will suppose, throughout this paper, that the neutralino relic density was produced in the ``standard'' thermal way, i.e. under the
assumptions that neutralinos were produced thermally thanks to their interactions with other particles in the primordial plasma, and that they decoupled while the Universe was radiation-dominated. Then, their present relic density is given by \cite{Gondolo:1990dk}
\be
\label{Omegathermal}
\Omega_{\rm DM} h^2 =\frac{8.7\times 10^{-11} \ {\rm GeV}^{-2}}{\sqrt{g*}\int_{x_f}^{\infty}\langle \sigma_{\rm ann}v\rangle x^{-2}} \ ,
\ee
where the $g*$ parameter accounts for the number of degrees of freedom at freeze-out, $x\equiv m/T$, i.e. temperature over mass, and the subscript $f$ denotes the freeze-out time, $T_f\simeq m/20$.  Besides, $\langle \sigma_{\rm ann}v\rangle$ stands for the thermal-averaged annihilation cross section (times the velocity).  Thus, in order to reproduce the observed relic density (\ref{OmegaDM}) the neutralinos must annihilate at early times with a suitable cross section.

From the naturalness point of view, an interesting case occurs when $\chi_1^0$
is close to a pure state. Then, roughly speaking, $\sigma \propto
m_{\chi_1^0}^{-2}$ and therefore, in order to reproduce (\ref{OmegaDM}), there
is in principle no need of any fine-arrangement of the parameters in the
${\cal M}_{\chi^0}$ matrix; only a particular value of $m_{\chi_1^0}$,
i.e. $\sim$ $M_1$, $M_2$ or $\mu$, depending on the character of
$\chi_1^0$. Actually, the case of (close to) pure bino does not work, since its
annihilation rate in the early Universe is typically too small for any value
of $M_1$, leading to an overproduction of dark matter, totally inconsistent
with eq.~(\ref{OmegaDM}).
In contrast, the cases of (essentially) pure Higgsino or pure wino lead to the
correct relic density if their masses are, respectively, $\mu\simeq 1$ TeV or
$M_2\simeq 3$ TeV.\footnote{ It is not clear at the moment if the pure-wino case is
consistent with DM indirect detection \cite{Cohen:2013ama, Fan:2013faa,
  Hryczuk:2014hpa}, due to the large uncertainties involved.}

Notice that both cases lead to a rather heavy supersymmetric spectrum, which has two problems. First of all, the expectations to discover supersymmetry at the LHC decrease (actually, for the wino-LSP they vanish). Second, the heavier the spectrum, the more fine-tuned the model with respect to the EW breaking. 
It is therefore of interest to consider mechanisms that allow for lighter neutralinos, keeping a correct relic density.  
This can be achieved, provided that $\chi_1^0$ is mostly bino, or at least it possesses a substantial bino-component, and that there is an additional mechanism to increase $\langle \sigma_{\rm ann}v\rangle$.
There are three of such mechanisms, which have been extensively studied in the literature:

\begin{description}

\item[{\em i)}]  \hspace{0.3cm}Well-tempered neutralinos. If the parameters of the ${\cal M}_{\chi^0}$ matrix are finely chosen, $\chi_1^0$ may be a well-tempered neutralino \cite{ArkaniHamed:2006mb}, i.e. an appropriate mixture of bino and Higgsino (or bino, Higgsino and wino), such that it annihilates in the right amount at early times. Since the  $\propto M_Z$ off-diagonal entries in ${\cal M}_{\chi^0}$ are typically much smaller than $M_1,M_2$ and $\mu$, a significant mixing requires some of the latter parameters to be near-degenerate.

\item[{\em ii)}] \hspace{0.2cm} Funnels. If $\chi_1^0$ is close to a bino, it can annihilate resonantly via  $Z-$funnel, Higgs-funnel or $A-$funnel, provided its mass is nearly half of the mass of the funnel-particle.

\item[{\em iii)}]  \hspace{0.1cm}Co-annihilation.  The effective $\langle \sigma_{\rm ann}v\rangle$ increases if $\chi_1^0$ can co-annihilate with other fast-annihilating particle (e.g. a stop, a stau or a gluino). This requires their masses to be nearly-degenerate.

\end{description} 

In all the above cases, one can foresee the need of cancellations or delicate balances, and thereby fine-tuning.

The aim of this paper is to analyze all these possibilities in detail, evaluating the associated fine-tuning. In some cases, this requires to re-visit the concept of fine-tuning itself, because the extrapolation of  the ``standard criterion", eq.~(\ref{BG}), to the relic density is not always appropriate. 
The paper is organized as follows. In section \ref{sec:FT}, we will review the different measurements of the fine-tuning. 
Sections \ref{sec:binoHiggs}, \ref{sec:binowino}, \ref{sec:funnels} and \ref{sec:coann} are devoted to the different scenarios for DM within the MSSM. 
In section \ref{sec:EWFT}, we make a connection between the fine-tuning in DM and the electroweak fine-tuning. 
Section \ref{sec:HiggsinoDM} is devoted to accommodating Higgsino DM in the MSSM and finally our conclusions are presented in section \ref{sec:conclu}.

\section{The measure of the fine-tuning}
\label{sec:FT}
In the few places of the literature where the fine-tuning associated with the DM relic density has been considered, the criterion to quantify it has always been the standard one, i.e. a direct extrapolation of the EW fine-tuning criterion (\ref{BG}) replacing $v^2$ by $\Omega_{\rm DM}$,
\be
\label{BGDM}
\Delta^{\rm (DM)}_i = \frac{d \log \Omega_{\rm DM}}{d \log \theta_i}\ ;\ \ \ \ \ \ \ \Delta^{\rm (DM)}\equiv\max\left\{\Delta^{\rm (DM)}_i\right\}.
\ee

However, behind this ``standard measure" there are implicit assumptions (seldom stated). If these assumptions do not hold, then the standard criterion may be misleading. In the next subsection, we compile those assumptions, and later we will show instances where those conditions are not fulfilled and therefore the standard criterion is not applicable. As we will see in the following sections, these instances are actually realized in several cases of DM production, which requires to improve the criterion to quantify the fine-tuning.

\subsection{Assumptions behind the standard fine-tuning criterion}
\label{subsec:assump}
Let us now analyze the statistical meaning of the standard fine-tuning criterion, eq.~(\ref{BG}). For the sake of simplicity, we will consider a single and representative $\theta-$parameter, e.g. the one producing the maximum $\Delta$ (usually $\theta$ is a soft mass or the $\mu-$parameter),
\bea
\label{BG2}
\Delta_{\theta}=\frac{\partial \log v^2}{\partial \log \theta}\  .
\eea
\begin{figure}[ht]
  \centering
    \includegraphics[width=8.5cm]{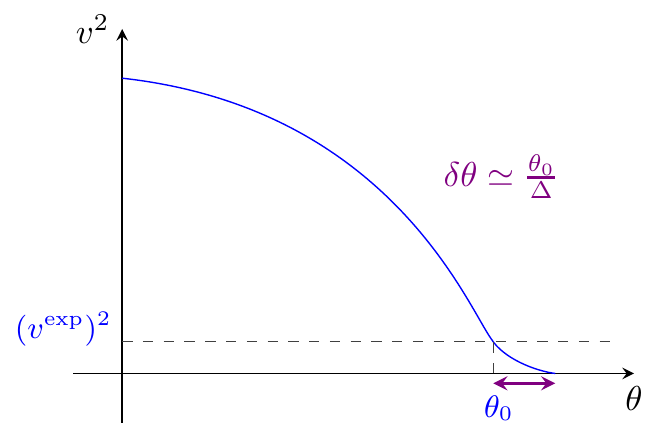}
  \caption{Schematic representation of the statistical interpretation of the standard fine-tuning criterion as the (inverse of the) $p-$value, $\Delta^{-1} = \delta\theta/\theta_0$.}
  \label{fig:pepe}
\end{figure} 

As it is known, the issue of the value of $v^2$ is that it receives contributions of the size of the soft squared-masses, which are typically ${\cal O}(100)$ times larger; thus a somewhat artificial cancellation among these contributions is required. 
Since for non-tuned values of the soft terms (represented by $\theta$), $v^2$ tends to be too large, one can estimate the small range of $\theta$ for which $v^2$ is abnormally small, say $v^2\lsim (v^{\rm exp})^2$. 
Expanding $v^2(\theta)$ at first order around the value $\theta^0$, which gives $(v^{\rm exp})^2$, $v^2(\theta_0+\delta\theta)\simeq v^2(\theta_0) + (\partial v^2(\theta)/\partial \theta)_{\theta_0}\ \delta\theta$,
we find that only for a small neighbourhood $\delta \theta\simeq \theta^0/\Delta_{\theta}$ around this point, $v^2$  is equal or smaller than the experimental value (see Fig.~\ref{fig:pepe}).
Therefore, if one assumes that $\theta$ could reasonably have taken any value of the order of magnitude of $\theta^0$, then only for a small fraction $\sim \left|{\delta \theta}/{\theta^0}\right| \simeq \Delta_{\theta}^{-1}$ of the $\theta$ values one gets $v^2\lsim (v^{\rm exp})^2$; this is the rough probabilistic meaning of $\Delta_{\theta}$ \cite{Ciafaloni:1996zh, Casas:2014eca}.
Consequently, $\Delta$ can be interpreted as the inverse of the $p$-value to get $v^2$ equal to the observed value or even smaller,
\begin{eqnarray}
\label{toy2}
p{\rm -value} \simeq \left|\frac{\delta \theta}{\theta_0}\right| \equiv \Delta^{-1}\ .
\end{eqnarray}

Then, we can summarize the implicit assumptions behind the standard fine-tuning criterion, eq.~(\ref{BG}):

\begin{enumerate}

\item The possible values of a $\theta-$parameter are distributed, with approximately flat probability, in the $\sim [0,\theta^0]$ range (flat prior in the Bayesian language). Note that, in fact, this represents two assumptions.

\item The expansion of $v^2(\theta)$ at first order captures its behaviour in the neighbourhood of interest.

\end{enumerate}
If any of these assumptions is not fulfilled, then the standard criterion has to be re-visited. Before showing some typical examples where this can happen, let us add some comments on the above conditions.

The assumed range for $\theta$ does not need to be $[0, \theta_0]$, any range of the same length, e.g. $[\theta_0/2, 3\theta_0/2]$, works equally well. The idea is that the range for $\theta$ should be of the same order than its actual value, $\theta_0$, so that the latter is a typical value. It could be argued that in the upper half of the previous alternative range, i.e. $[\theta_0, 3\theta_0/2]$ it happens that $v^2\leq (v^{\rm exp})^2$, simply because $v^2=0$ for most of it. Then the $p-$value would be $\simeq 1/2$. Nevertheless, the region where $v^2$ is strictly vanishing should not be counted since it does not represent any extreme case but simply the case where the Higgs mass-squared parameter is positive. An equivalent way to take this fact into account is to directly define the $p-$value for the mass parameter itself, $m^2$, instead of $v^2$ (both are related by $v^2=-m^2/\lambda$). Then, one evaluates the probability of having $|m^2|\leq |m^{\rm exp}|^2$, giving a similar result as eq.~(\ref{toy2}).

The previous discussion illustrates the fact that there is always an $\sim {\cal O}(1)$ factor of arbitrariness for the fine-tuning measure. E.g. choosing the range of $\theta$ two times longer than the previous one increases $\Delta_\theta$ by a factor 2.

It is also worth mentioning that the standard fine-tuning criterion is also valid for alternative choices of the prior and range of $\theta$. E.g., if one assumes that $\theta$ has a logarithmic prior, i.e. its a-priori probability distribution is flat in the logarithm, ${\cal P}(\theta) \propto 1/\theta$, then a similar argument leads to the same eq.~(\ref{toy2}), provided that the range of $\theta$ satisfies $\log|\theta^{\rm max}/\theta^{\rm min}|=1$.

Let us finally mention that the previous discussion about the statistical meaning of the fine-tuning can be expressed in Bayesian terms, following a Bayesian analysis of the probability distribution in the parameter space, see refs.~\cite{Cabrera:2008tj,Cabrera:2009dm}.

\subsection{Examples}

The EW fine-tuning stemming from the artificial cancellation between different contributions in order to get $v^2$ small enough, does reasonably fulfil conditions 1 and 2 of the previous subsection. In other words, in the MSSM the dependence of $v^2$ on the relevant soft terms and the $\mu-$parameter goes as in Fig.~\ref{fig:pepe} or behaves in a similar manner. So the standard criterion to quantify the EW fine-tuning is sound. 

\begin{figure}[ht]
  \centering
      \includegraphics[width=8cm]{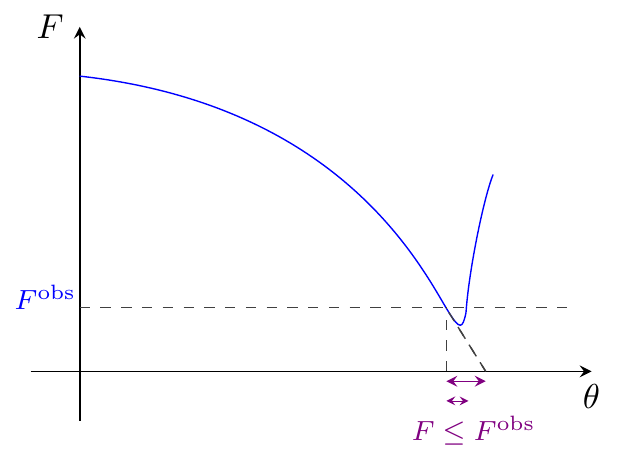}
  \caption{A hypothetical case where the standard criterion (see Fig.~\ref{fig:pepe}) underestimates the severity of the fine-tuning.
  }
  \label{fig:standard_criterion_1}
\end{figure}

\begin{figure}[ht]
  \centering
      \includegraphics[width=9cm]{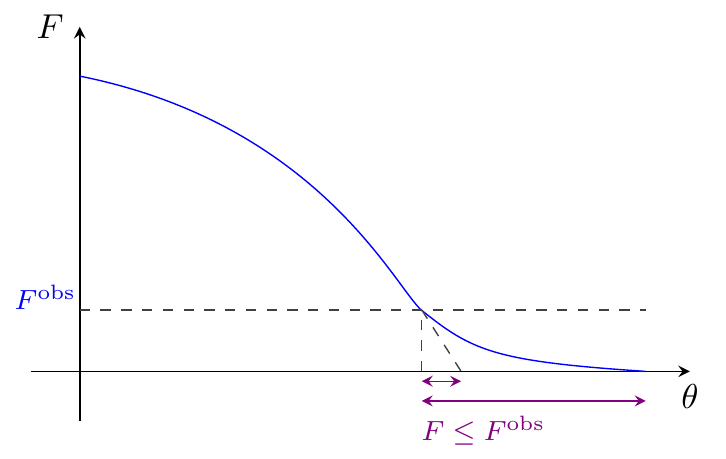}
  \caption{A hypothetical case where the standard criterion (see Fig.~\ref{fig:pepe}) overestimates the severity of the fine-tuning.}
  \label{fig:standard_criterion_2}
\end{figure}

Now, let us suppose that the fine-tuned quantity, say $F$, has a different dependence on $\theta$. Figs.~\ref{fig:standard_criterion_1} and \ref{fig:standard_criterion_2} show two instances in which this happens in distinct ways. In Fig.~\ref{fig:standard_criterion_1}, the hypothetical $F$-quantity acquires its experimentally observed value $F^{\rm (obs)}$ for some value $\theta_0$. 
However,  there is no value of $\theta$  for which $F$ vanishes.
Hence, the region of $\delta \theta$ for which $F\leq F^{\rm (obs)}$ cannot be approximated by $\delta \theta\simeq \theta_0/\Delta_{\theta}$. 
The actual $\delta \theta$ region is narrower and thus the actual $p-$value is smaller and the fine-tuning is more severe.  This example also illustrates another potential departure from the conditions 1 and 2 stated in the previous subsection. 
Obviously, if the value of $\theta_0$ that reproduces $F^{\rm (obs)}$ lies {\em very} close to the minimum of the $F(\theta)$ function, then the fine-tuning is enormous, since essentially the $\delta \theta$ region  for which $F\leq F^{\rm (obs)}$ shrinks to a point. Nevertheless, a blind application of the standard criterion would lead to $\Delta\rightarrow \infty$. Evidently, the problem is that in this case $\theta_0$ would be a stationary point and thereby it would be no longer justified to truncate the expansion at first order (condition 2 in the previous subsection). 
An important lesson is that sensitivity is not always equivalent to fine-tuning, and sometimes the measure of sensitivity, which is what the standard criterion provides, does not reflect the actual degree of fine-tuning. As we will see, when the relic density gets the observed value thanks to the annihilation of neutralinos through $Z$, Higgs or $A$ funnels, $\Omega_{\rm DM}$ has a dependence on the MSSM parameters similar to that of Fig.~\ref{fig:standard_criterion_1}.

Fig.~\ref{fig:standard_criterion_2} shows another example in which the assumptions for the applicability of the standard fine-tuning criterion do not hold. In this case, the truncation of $F(\theta)$ at first order is not good enough to evaluate the region $\delta\theta_0$ for which $F\leq F^{\rm (obs)}$. Here, the linear approximation leads to an underestimation of $\delta\theta_0$, so that the actual $p-$value is larger and the fine-tuning is less severe than the one obtained from the standard criterion. As we will see, the example of Fig.~\ref{fig:standard_criterion_2} describes schematically the dependence of $\Omega_{\rm DM}$ on the MSSM parameters when the DM is wiped out through co-annihilations.

\section{Well tempered bino-Higgsino}
\label{sec:binoHiggs}
Consider first the well-tempered Higgsino/bino, i.e. the case in which the lightest neutralino is a combination of bino and Higgsino. Obviously, this scenario includes the pure-Higgsino case as a particular and important limit (recall that, in contrast, the pure-bino limit is not viable unless additional mechanisms for DM annihilation are present). As mentioned in the introduction, the appeal of this setup is that it enables cases where the LSP is lighter than in the pure-Higgsino case, since the annihilation of LSPs becomes reduced thanks to the bino component. 
On the other hand, the possibility to find DM in (spin-independent) direct detection experiments through 
the neutralino elastic scattering off quarks
 mediated by a Higgs boson, is also higher,
due to the Higgsino-bino-Higgs coupling. Indeed, present bounds on direct
detection are able to exclude a large portion of the
bino-Higgsino parameter space \cite{Cheung:2012qy,Crivellin:2015bva}. However, it
still remains as an interesting scenario, with relevant implications for the
LHC and DM direct detection searches. It is also an illustrative example of the
subtleties involved in the calculation of the DM fine-tuning.

From the four parameters that define the neutralino mass matrix, eq.~(\ref{M_X}),
the most relevant ones here are $M_1$ and $\mu$. $M_2$ plays a negligible
role, unless it happens to be quite degenerate with $M_1$ and $\mu$, in which
case the neutral and charged wino would contribute to DM co-annihilation processes.
This would correspond to the bino/wino/Higgsino scenario, to be analyzed in
the next subsection. Consequently, for the bino-Higgsino analysis, $M_2$ can
be made large enough for winos to be ignored. In addition, as stated in the
introduction, one needs $\tan \beta$ at least moderately large, say
$\tan\beta\gsim 6$, in order to maximize the tree-level Higgs mass ($m_h^{\rm
  tree}\leq M_Z$). In this way, we avoid the necessity of large radiative
corrections to increase $m_h$ up to its experimental value, which would
require enormous stop masses and thereby an extremely large EW fine-tuning. Since the
aim of this work is to explore as less fine-tuned as possible supersymmetric
DM, we will ignore the small $\tan\beta$ regime. On the other hand, in the
large-$\tan\beta$ regime the precise value of $\tan\beta$ is not very
important, because it hardly affects the numerical values of the ${\cal
  M}_{\chi^0}$ entries. In conclusion, concerning the potential fine-tuning to
arrange the correct DM relic density, $\tan\beta$ can be safely ignored.

\begin{figure}[ht]
  \centering
  \includegraphics[width=10cm]{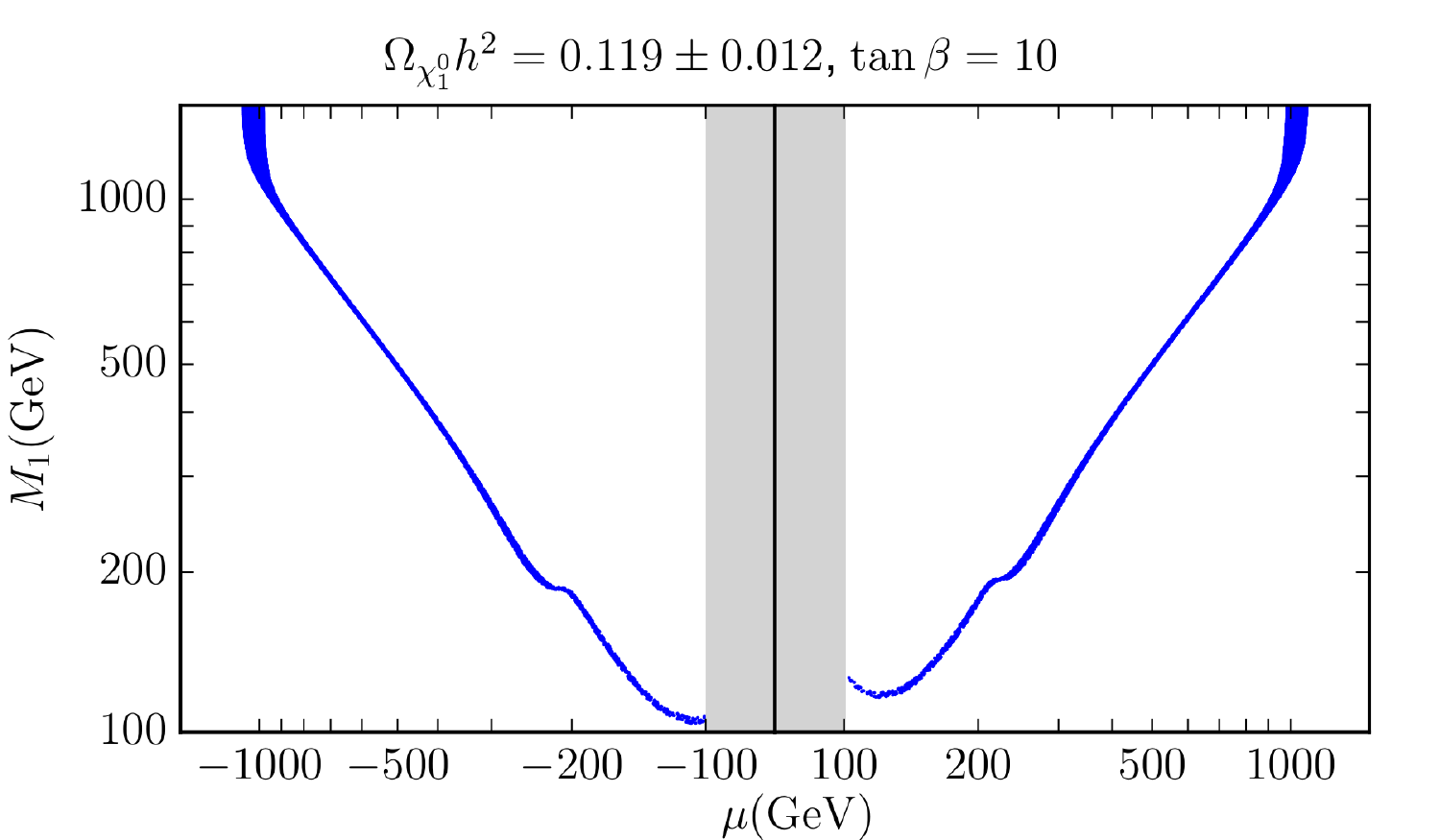}
  \caption{Region of the $\mu-M_1$ plane that leads to the observed DM relic density, $\Omega_{\chi_1^0}h^2=0.119\pm 0.012$, in a well-tempered bino-Higgsino scenario (blue bands). The grey band is excluded by LEP limits on charginos.}
  \label{fig:binohino_mu-M1}
\end{figure} 

\begin{figure}[ht]
  \centering
  \hspace{-1cm}
  \includegraphics[width=.52\linewidth]{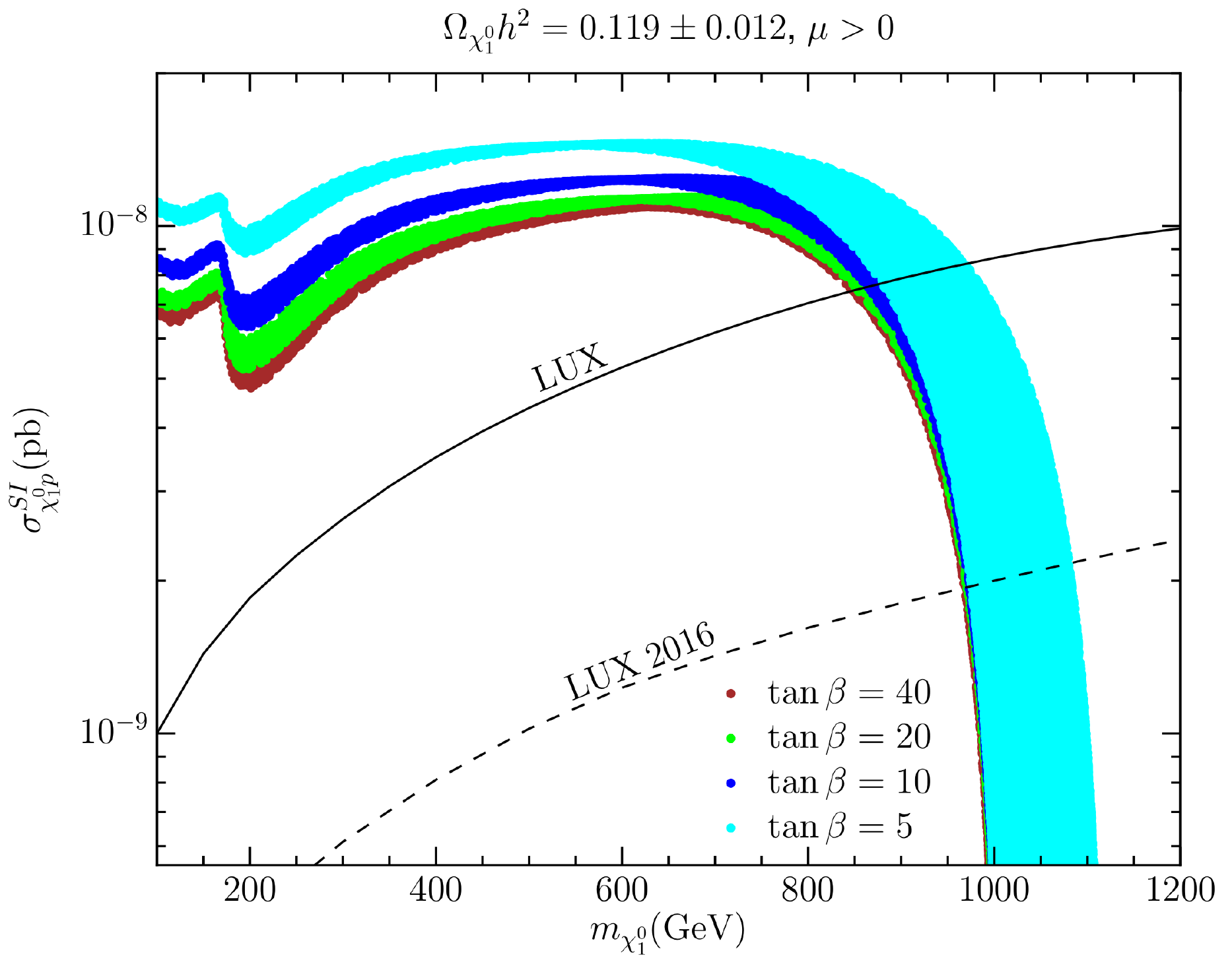}
  \includegraphics[width=.52\linewidth]{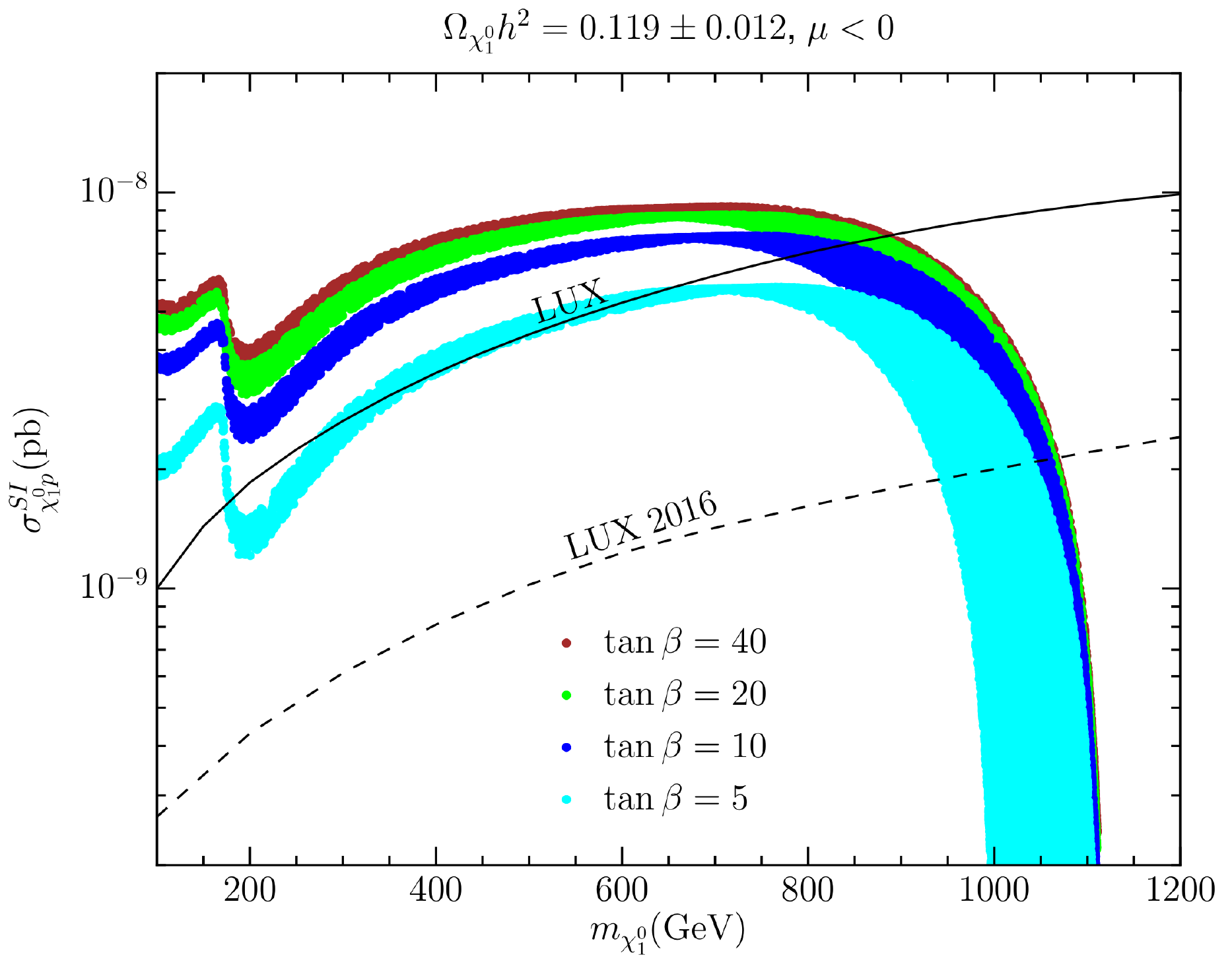}
  \caption{Spin-independent neutralino-proton cross section for the bino-Higgsino scenario for $\mu>0$ (left) and $\mu<0$ (right) and different values of $\tan \beta$. 
  The current exclusion line and the preliminary 2016 limit from LUX (assuming that the neutralino is entirely made of bino-Higgsinos) are shown as solid and dashed lines, respectively.  
  The XENON 1T and LZ projected sensitivities lie below the horizontal axes. }
  \label{fig:binohino_Mn1-sSI}
\end{figure}

Fig.~\ref{fig:binohino_mu-M1} shows in blue the region in the $\mu-M_1$
plane where $\Omega_{\chi_1^0}h^2=0.119\pm 0.012$. It is located close to the
$|\mu|=M_1$ lines, something required in order to get a non-trivial
bino-Higgsino mixture. The calculation has been performed using
  \texttt{SOFTSUSY-3.6.2} \cite{Allanach:2001kg} to compute the mass spectrum,
  \texttt{micrOMEGAs-4.1.8} \cite{Belanger:2014vza,Belanger:2013oya} for the relic
  density and direct detection cross section, and \texttt{MultiNest-3.9}
  \cite{Feroz:2007kg,Feroz:2008xx,Feroz:2013hea} to efficiently 
  explore the parameter space. The current LUX exclusion line \cite{Akerib:2015rjg} and the preliminary LUX 2016 limit \cite{Manalaysay:2016} for the two
signs of $\mu$ are presented in Fig.~\ref{fig:binohino_Mn1-sSI}, showing the
impressive power of present and future experiments of DM direct detection to
exclude large regions of the parameter space. In fact, the (non-visible)
XENON 1T and LZ projected sensitivities lie below the horizontal axes, so that they will
potentially probe the whole scenario \cite{XENON1T,Aprile:2015uzo,Cushman:2013zza}. 

Let us now consider the DM fine-tuning issue. From Fig.~\ref{fig:binohino_mu-M1} it is clear that a certain fine-tuning is required for the viability of the model, since the (low-energy) values of $|\mu|$ and $M_1$ must be quite degenerate.  In the absence of a theoretical argument to justify such coincidence, this clearly represents a fine-tuning. 

Before attempting to quantify it, let us mention an interesting and fortunate
fact. The degree of naturalness of a physical scenario must be evaluated by
examining the behaviour of the fine-tuned quantities with respect to the
independent parameters of the theory, see e.g. the standard measure of
eq.~(\ref{BG}). Here, ``independent'' means that there is no known theoretical
connection between them (or no connection based on some
specific model is assumed). For the present case, the relevant independent parameters are
the initial (high-energy) values of the soft parameters and
$\mu$. E.g. $\tan\beta$ is a derived parameter, which depends on the initial
ones in a complicated way. Nevertheless, as mentioned above, the dependence of
${\cal M}_{\chi^0}$ on $\tan\beta$ is very weak, so we can ignore its impact
on the fine-tuning. Now, the fortunate fact is that the remaining three
relevant (low-energy) parameters, involved in ${\cal M}_{\chi^0}$, namely $M_1,
M_2$ and $\mu$, are essentially in one-to-one multiplicative correspondence with
the three initial (high-energy) parameters,
\begin{eqnarray}
\label{Rs}
\left.M_i\right|_{LE}&=&c_{M_i}\left.M_i\right|_{HE}, \quad i=1, 2, \nonumber  \\  
\left.\mu\right|_{LE}&=&c_\mu\left.\mu\right|_{HE} \ ,
\end{eqnarray}
where the HE (LE) subscript denotes high- (low-) energy, and the values of the
$c-$coefficients depend on the value of the HE scale (see
ref.~\cite{Casas:2014eca} for a recent computation). However, for fine-tuning
purposes the particular values of the $c$'s, and thus the choice of the HE
scale, are irrelevant. E.g. for the standard 
fine-tuning measure, eq.~(\ref{BGDM}), the logarithmic derivatives are the
same evaluated with respect to the HE or the LE parameters. 
This fact simplifies life considerably and allows to work just with the
low-energy parameters, producing results on $\Delta^{\rm (DM)}$ which are pretty
general, in particular $\Delta^{\rm (DM)}$ is essentially independent of the HE
scale and the values of the remaining MSSM parameters, which is
remarkable. Incidentally, this is not the case for the EW fine-tuning, where a
specific analysis must be performed for each model.

\begin{figure}[ht]
  \centering
  $$\includegraphics[width=10cm]{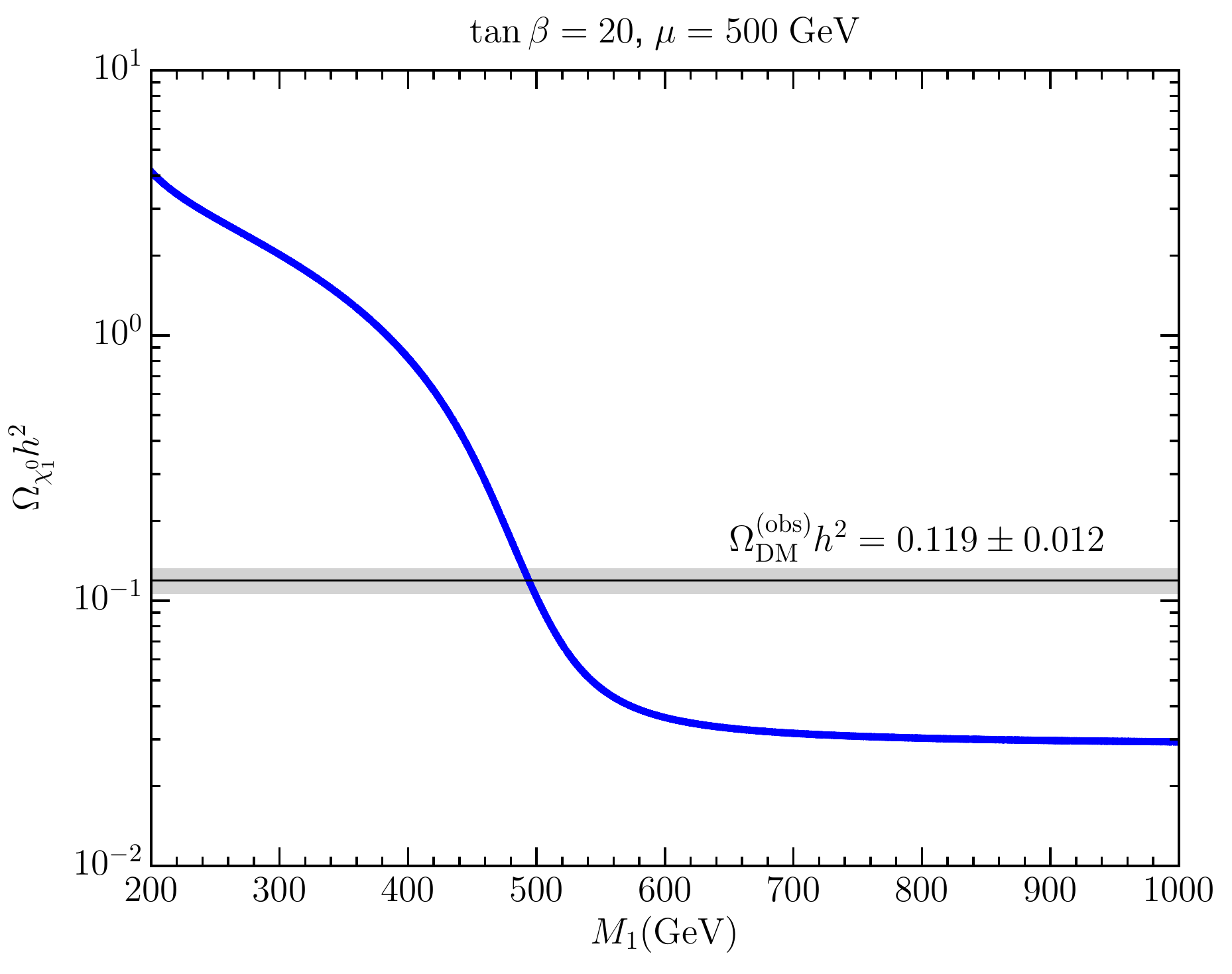} $$   
  \caption{$\Omega_{\chi_1^0} h^2$ vs $M_1$ in the well-tempered bino-Higgsino scenario for fixed values of $\mu$ and $\tan\beta$.}
  \label{fig:binohino_mu500_M1-Omega}
\end{figure}

Let us now compute the DM fine-tuning. Before relying on the standard measure, eq.~(\ref{BGDM}), it is convenient to test if the conditions 1 and 2 listed in subsection \ref{subsec:assump} are fulfilled. 
In other words, we should check the dependence of $\Omega_{\chi_1^0}$ on the $\mu$ and $M_1$ parameters (the only relevant ones for this scenario). Since the tuning is precisely between these two parameters, it is enough to consider one of them, say $M_1$.\footnote{This has the advantage of avoiding interference with the EW fine-tuning, for which $\mu$ is a very relevant parameter, unlike $M_1$.} Fig.~\ref{fig:binohino_mu500_M1-Omega} shows such dependence for a fixed value of $\mu$. 

As expected, only for a small interval of $M_1$, $\Omega_{\chi_1^0}$ is consistent with the observed value. Nevertheless, concerning the fine-tuning, the important issue is that, {\em typically}, $\Omega_{\chi_1^0}$ is much larger or much smaller than $\Omega_{\rm DM}^{\rm (obs)}$. It requires a tuning between $M_1$ and $\mu$ for $\Omega_{\chi_1^0}$ to be in the vicinity of the observed value. Now, if we consider that the range of $M_1$ is $[0, M_1^{(0)}]$, where $M_1^{(0)}$ is the value that reproduces $\Omega_{\rm DM}^{\rm (obs)}$, then the standard measure of eq.~(\ref{BGDM}) and its interpretation in terms of $p-$value, i.e. the probability of getting $\Omega_{\chi_1^0}\leq\Omega_{\rm DM}^{\rm (obs)}$, is justified. 
However, changing the limits of the range to e.g. $[{M_1^{(0)}}/{2}, \ {3M_1^{(0)}}/{2}]$ jeopardizes the $p-$value interpretation, because there is a large interval of $M_1$ for which $\Omega_{\chi_1^0}\leq\Omega_{\rm DM}^{\rm (obs)}$. A way out to this difficulty is to change the definition of the fine-tuned quantity. 
Instead of $\Omega_{\chi_1^0}$, we can use the mixing angle, $\theta$, between the bino and the Higgsino. More precisely, upon diagonalization of ${\cal M}_{\chi^0}$, given by  eq.~(\ref{M_X}), one gets

\begin{eqnarray}
\label{tan2theta}
|\tan2\theta|\simeq\frac{\sqrt{2}\ s_WM_Z}{|\mu-M_1|}
 \ .
\end{eqnarray}

It is worth noting that $\theta$ is a physical quantity, in direct correspondence with $\Omega_{\chi_1^0}$, which could have been experimentally measured before $\Omega_{\rm DM}^{\rm (obs)}$. If $\tan2\theta$ is large, this clearly denotes a fine-tuning between $M_1$ and $\mu$ in eq.~(\ref{tan2theta}). In terms of $\tan2\theta$ the $p-$value interpretation of the fine-tuning is much more transparent and robust than before: it is the probability of getting $|\tan2\theta|\geq |\tan2\theta^{\rm (obs)}|$. Assuming, as usual, a flat prior for $M_1$ in the region of interest, such $p-$value is simply
\begin{eqnarray}
\label{thetapvalue}
p-{\rm value}\ = \ \frac{2|\mu-M_1|}{|M_1|}
\ ,
\end{eqnarray}
independent of the position of the $M_1-$range limits.

\begin{figure}[ht]
  \centering
  $$\includegraphics[width=10cm]{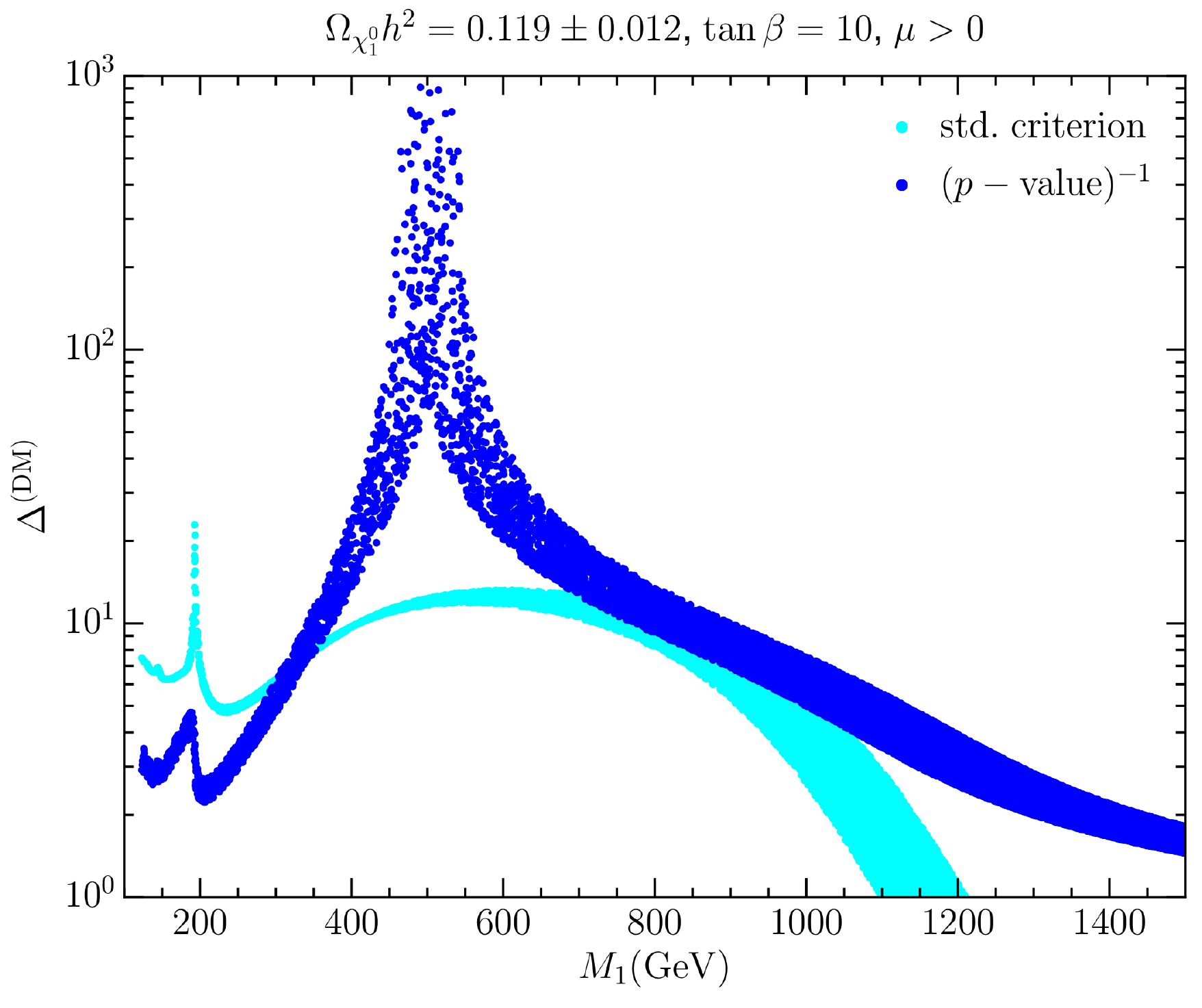} $$   
  \caption{Fine-tuning in the well-tempered bino-Higgsino scenario for $\mu>0$, calculated using the ``standard criterion" [see eq.~(\ref{BGDM})] (cyan band) and using the $p-$value criterion [i.e., the inverse of the $p-$value evaluated as in eq.~(\ref{thetapvalue})] (blue band). The width of the bands corresponds to the uncertainty in the relic density, $\Omega_{\chi_1^0}h^2=0.119\pm 0.012$.}
  \label{fig:binohino_M1-FT}
\end{figure}

Fig.~\ref{fig:binohino_M1-FT} shows the fine-tuning calculated with the standard criterion eq.~(\ref{BGDM}) and the one estimated by the inverse of the $p-$value, eq.~(\ref{thetapvalue}).\footnote{Let us note the funny fact that if one had applied the standard fine-tuning criterion to the physical quantity $\tan2\theta$ instead of $\Omega_{\chi_1^0}$, i.e. $\Delta=d\log \tan2\theta/d\log M_1$, the result  would have become essentially equivalent to
the inverse of the $p-$value, eq.~(\ref{thetapvalue}). This shows that the standard criterion is not always robust under changes in the definition of the fine-tuned quantity. However, starting directly with the $p-$value criterion is much more trustworthy.} Needless to say, a $(p-{\rm value})^{-1} = {\cal O}(1)$ is completely normal for a non-fine-tuned quantity, so fine-tunings below 5 or even 10 are not significant.

Qualitatively both criteria give similar results. In particular, the region around $M_1=500-600$ GeV is the most fine-tuned one, since it is the one that requires $\mu=M_1$ with more precision (it corresponds to maximal bino-Higgsino mixing angle). This can also be seen at naked eye in Fig.~\ref{fig:binohino_mu-M1}, by examining the width of the $\Omega_{\chi_1^0}h^2=0.119\pm 0.012$ (blue) band, which narrows in that region. Quantitatively, the fine-tuning estimated by the $p-$value criterion is in general more severe and, in our opinion, more reliable for the above-discussed reasons. Interestingly, for the $M_1\gsim 950$ GeV region, which is the one allowed by LUX, see Fig.~\ref{fig:binohino_Mn1-sSI}, the tuning is rather small, even non-significant. This includes, of course, the $M_1>\mu\simeq 1$ TeV region, for which the lightest neutralino is essentially a Higgsino. Actually, in this limit the precise value of $M_1$ is irrelevant, and the dependence of $\Omega_{\chi_1^0}$ on $\mu$, namely $\Omega_{\chi_1^0}\propto \mu^2$, does not entail any fine-tuning, as expected, see the discussion in section~\ref{sec:intro}.

\section{Well-tempered bino-wino(-Higgsino)}
\label{sec:binowino}
By inspection of the ${\cal M}_{\chi^0}$ mass matrix, eq.~(\ref{M_X}), it is clear
that a substantial bino-wino mixing requires a not too-large $\mu$. Thus,
assuming again moderate or large $\tan\beta$, this scenario has three relevant
parameters, $M_1, M_2$ and $\mu$. Furthermore, the Higgsino gets also mixed, so that 
the scenario really becomes well-tempered bino-wino-Higgsino.

\begin{figure}[ht]
  \centering
  $$\includegraphics[width=10cm]{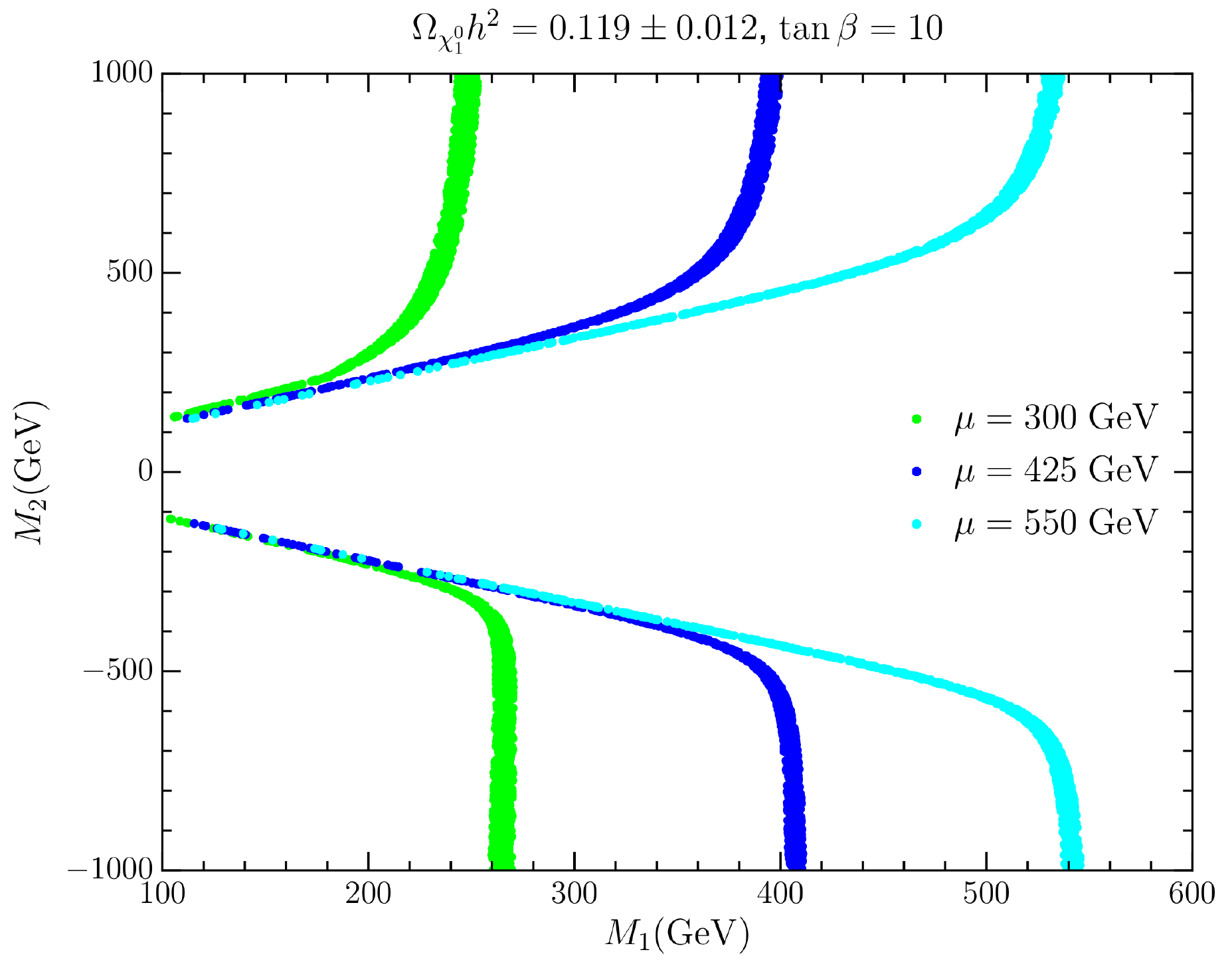} $$   
  \caption{Region of the $M_1-M_2$ plane that leads to the observed DM relic density, $\Omega_{\chi_1^0}h^2=0.119\pm 0.012$, for three different values of $\mu$ in a well-tempered neutralino scenario.}
  \label{fig:binowinohino_M1-M2}
\end{figure}

However, there is a special and physically relevant limit, where things
become simpler. Namely, for large enough $\mu$, the mixing between bino and
wino (and Higgsino) is small. In that regime, provided $M_1$ and $M_2$ are
nearly-degenerate, the neutralino annihilation is dominated by co-annihilation
with winos (more precisely, by wino-annihilation provided these are in thermal equilibrium with the lightest neutralino)
 \cite{ArkaniHamed:2006mb} and is almost independent of the value of
$\mu$.  All this is illustrated in Fig.~\ref{fig:binowinohino_M1-M2}, which
shows the region in the $M_1-M_2$ plane where $\Omega_{\chi_1^0}h^2=0.119\pm
0.012$ for three different values of $\mu$ and the two signs of $M_2$. For
$|\mu|\gg |M_1|, |M_2|$ the solution is close to the straight band
$|M_1|\simeq |M_2|$ and is quite independent of $\mu$ (the larger $\mu$, the
more independent the solution). This scenario can still be called well-tempered bino-wino, even 
though $\chi_1^0$ is mostly bino.

In this figure, the $|\mu|\simeq |M_1|\ll |M_2|$ regions
(nearly vertical segments of the coloured bands) are also visible. They correspond to the bino-Higgsino solution
(analyzed in the previous section), and are quite independent of the value of
$|M_2|$. Likewise, the $|\mu|\sim|M_1|\sim |M_2|$ regions (short, curved parts of the bands) correspond to the bino-wino-Higgsino case,
to be discussed later.

For the fine-tuning discussion of the bino-wino scenario, it is useful  to consider some analytical approximations. Note that, since this is a co-annihilation scenario, the averaged annihilation cross-section $\langle \sigma_{\rm ann}v\rangle$ in eq.~(\ref{Omegathermal}) must be replaced by \cite{ArkaniHamed:2006mb}
\begin{eqnarray}
\label{sigma_coan}
\langle \sigma_{\rm eff}v\rangle=\frac{\sum_{i,j=1}^N w_iw_j \sigma_{ij} x^{-n}}{(\sum_{i=1}^N w_i^2)^2}\ , \ \ \ w_i=\left(\frac{m_i}{m_1}\right) ^{3/2} e^{-x\left(\frac{m_i}{m_1}-1\right)}\ ,
\end{eqnarray}
where $N$ is the number of co-annihilating species (in this case the bino and the three winos), $m_1$ is the lowest mass (in this case $\sim M_1$) and the $ij\rightarrow$ SM SM annihilation-cross-sections are parametrized as the dominant term in the velocity- (or equivalently $x$-) expansion
\begin{eqnarray}
\label{sigma_coan2}
\langle \sigma_{ij}v\rangle\simeq \sigma_{ij}x^{-n}
\ .
\end{eqnarray}

\begin{figure}[ht]
  \centering
  \hspace{-0.8cm}
  \includegraphics[width=.5\linewidth]{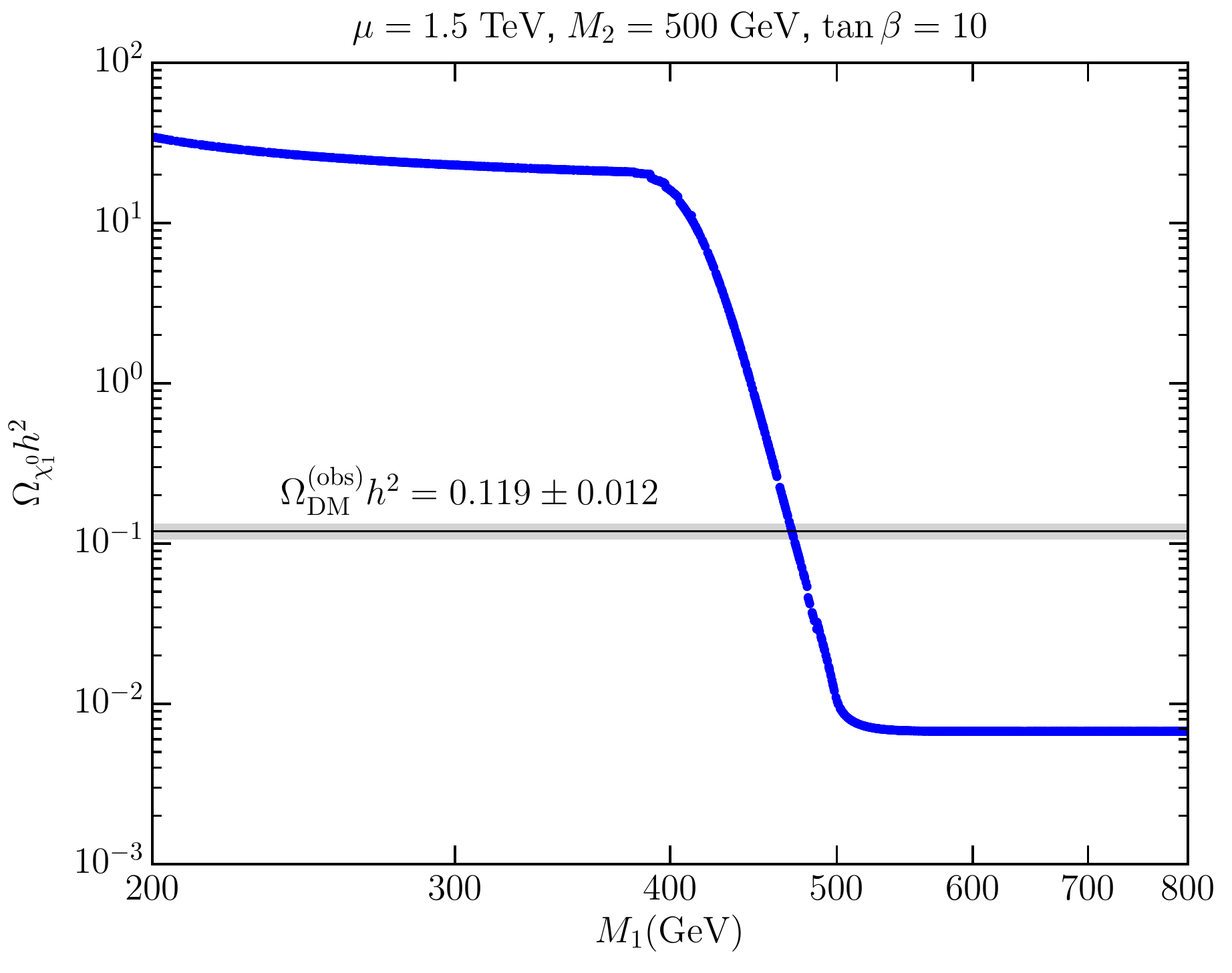}   
  \includegraphics[width=.4875\linewidth]{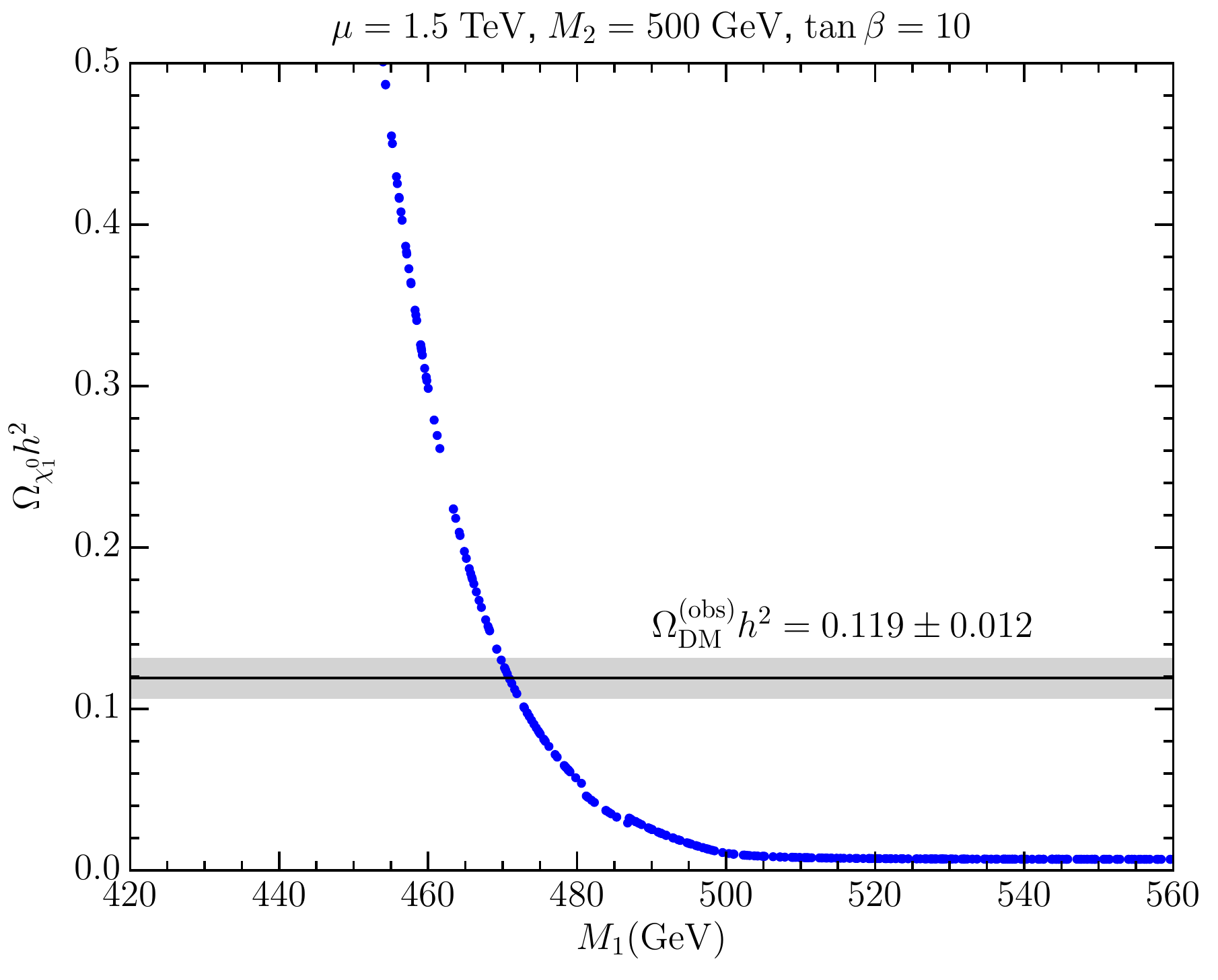} 
  \caption{$\Omega_{\chi^0_1} h^2$ vs $M_1$ in the well-tempered bino-wino scenario for fixed $\mu=1.5$ TeV, $M_2=500$ GeV and $\tan\beta=10$. Left (right) panel shows the relic abundance in logarithmic (linear) units.}
  \label{fig:binowino_M1-Omega}
\end{figure}

Under these circumstances the neutralino relic abundance is mostly determined by the $\tilde W$ annihilation processes, whose cross sections go as $\sim g^4/M_2$. Plugging numerical factors one arrives to a good approximate expression for the relic density \cite{ArkaniHamed:2006mb}, 
\begin{eqnarray}
\label{Omega_coan}
\Omega_{\chi^0_1} h^2\simeq 0.13\left(\frac{M_2}{2.5\ {\rm TeV}}\right)^2\frac{1}{R_{\tilde W}}
\ ,
\end{eqnarray}
where
\begin{eqnarray}
\label{RW}
R_{\tilde W}=\int_0^1dy \left[ 1+\frac{1}{3}\left(\frac{M_1}{M_2}\right)^{3/2} e^{\frac{x_f}{y}\left(\frac{M_2}{M_1}-1\right)}
\right]^{-2}\simeq \left(\frac{3}{4}\right)^2 e^{-\xi_{\tilde W}x_f \left(\frac{M_2}{M_1}-1\right)}
\ ,
\end{eqnarray}
with $\xi_{\tilde W}\simeq 1.7$. Recalling that $x_f\sim 20$, the previous equations (\ref{Omega_coan}, \ref{RW}) show a strong sensitivity of $\Omega_{\chi^0_1}$ to $M_1$. This is illustrated in Fig.~\ref{fig:binowino_M1-Omega}, which shows $\Omega_{\chi^0_1} h^2$ vs. $M_1$, using the complete numerical evaluation performed with \texttt{micrOMEGAs}, for fixed values of $M_2$ and $\tan\beta$. The value of $\mu$ is quite irrelevant provided is large enough ($\mu=1.5$ TeV in the figure).

From eqs.~(\ref{Omega_coan}, \ref{RW}) and Fig.~\ref{fig:binowino_M1-Omega}, one could foresee that the standard criterion will point to a severe fine-tuning. 
On the other hand, it should be noticed that the application of the standard recipe eq.~(\ref{BGDM}) to eqs.(\ref{Omega_coan}, \ref{RW}) leads to a value of the fine-tuning, $\Delta$, that is essentially independent of the mass difference 
$\Delta m= ||M_2|-|M_1||$, as $\Omega_{\chi_1^0}$ is dominated by the Boltzmann (exponential) factor in a substantial range of $M_1-$values. 
 This is counter-intuitive, since logically the fine-tuning should be more severe when $\Delta m$ is required to be smaller. 
 Fig.~\ref{fig:binowino_M1-Omega} (right panel), which shows the dependence of $\Omega_{\chi_1^0} h^2$ on $M_1$ in a linear scale, clarifies the connection of the standard fine-tuning measure to the $p-$value in this case.
Evidently, the exponential shape leads to an over-estimation of the fine-tuning when this is calculated with the standard criterion, compare Fig.~\ref{fig:binowino_M1-Omega} (right panel) to Fig.~\ref{fig:standard_criterion_2}. 
Again, we find that the simple ``$p-$value--like measure", $\Delta m/M_1$, offers a more sensible and robust description of the fine-tuning, as it happened in the  bino-Higgsino case analyzed in section~\ref{sec:binoHiggs}.

\begin{figure}[ht]
  \centering
  $$\includegraphics[width=10cm]{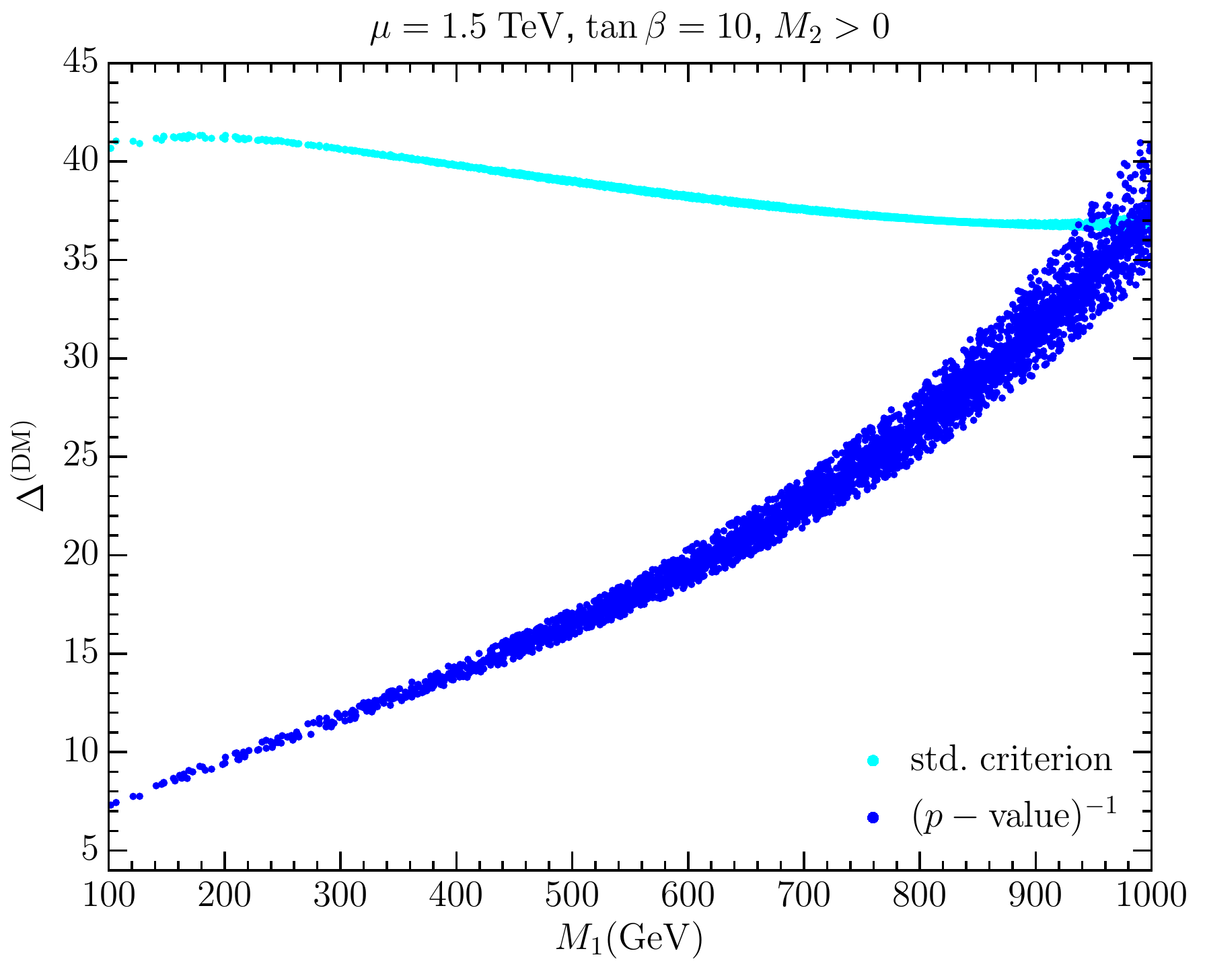} $$   
  \caption{Fine-tuning in the well-tempered bino-wino scenario for $\mu=1.5$ TeV, $\tan\beta=10$, calculated using the ``standard criterion" [see eq.~(\ref{BGDM})] (cyan band) and using the $p-$value criterion, i.e. $(\Delta m/M_1)^{-1}$ (blue band). The width of the bands corresponds to the uncertainty in the relic density, $\Omega_{\chi_1^0}h^2=0.119\pm 0.012$.}
  \label{fig:binowino_M1-FT}
\end{figure}

Fig.~\ref{fig:binowino_M1-FT} shows the performance of both criteria. For each value of $M_1$, the corresponding $M_2$ is chosen so that the observed relic density (\ref{OmegaDM}) is fulfilled (recall that the value of  $\mu$ is large and fairly irrelevant). 
Both $M_1$ and $M_2$ are defined at the $Q=M_1$ scale, and their values are close to the physical masses, $m_{\chi_1^0}$ and $m_{\chi_2^0},m_{\chi_1^\pm}$, respectively.
As discussed above, the standard criterion leads to an almost flat fine-tuning, independently of $M_1$ and $\Delta m$. The $p-$value criterion, however, varies considerably with $M_1$, showing a rather mild fine-tuning when the neutralino is light. 
The reason is that the heavier the wino, the less efficient its annihilation. Hence, in order to reproduce the relic density, the Boltzmann penalty in the co-annihilation process must be lessened, which requires a smaller $\Delta m/M_1$, and thus a higher fine-tuning.

\begin{figure}[ht]
  \centering
  \hspace{1cm}
  \includegraphics[width=10cm]{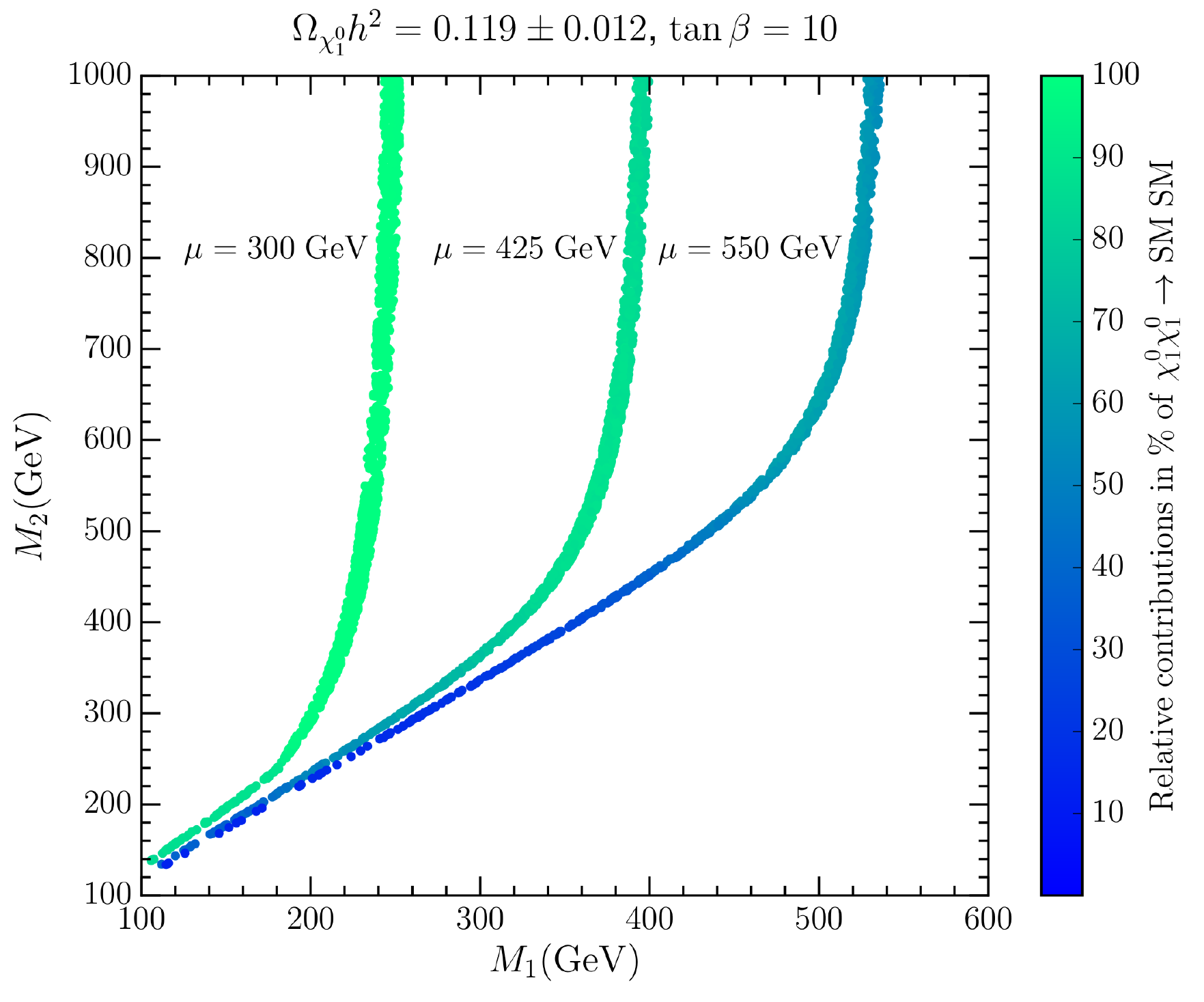} 
  \caption{As Fig.~\ref{fig:binowinohino_M1-M2}, zoomed on the positive region of the $M_1-M_2$ plane. The color code denotes the percentage of $\chi_1^0\chi_1^0\rightarrow {\rm SM\ SM}$ annihilation (the remaining DM annihilation proceeds mainly via co-annihilation with winos).}
  \label{fig:binowinohino_M1-M2p-Ann}
\end{figure}

Interestingly, the fine-tuning (evaluated with the $p-$value criterion) is milder when $|\mu|$ approaches the value of $|M_1|$ or $|M_2|$. 
This is due to the fact that as $|\mu|$ decreases the mixing between bino and wino increases. Then, the neutralino annihilation does not only occur through 
wino co-annihilation, as explained above, but also through direct $\chi_1^0\chi_1^0\rightarrow {\rm SM\ SM}$,  $\chi_1^0\chi_1^\pm\rightarrow {\rm SM\ SM}$ processes, thanks to the non-negligible wino component of the neutralino. 
Since the resulting annihilation is now more efficient, $|M_1|$ does not need to be that close to $|M_2|$. Consequently, the $p-$value is larger (and the fine-tuning less severe). 
This effect is illustrated in Fig.~\ref{fig:binowinohino_M1-M2p-Ann}, which displays the upper  part of Fig.~\ref{fig:binowinohino_M1-M2} (positive $M_1/M_2$ plane), but explicitly showing the percentage of $\chi_1^0\chi_1^0\rightarrow {\rm SM\ SM}$ annihilation.

 The situation depicted above connects with the bino-wino-Higgsino case, which occurs when the three relevant parameters, $\mu, M_1$ and $M_2$, have similar absolute values (curved segments of the bands in Fig.~\ref{fig:binowinohino_M1-M2p-Ann}). Intuitively, this case requires a more severe fine-tuning, as it requires a ``conspiracy" between three (a priori) independent parameters. It is therefore disfavoured from the point of view of naturalness, which is the main concern of this paper. One can try to estimate the related $p-$value. Assuming that $\mu$ is a given value, the separate $p-$values associated with the tuning of $M_1$ and $M_2$ are of order
\begin{eqnarray}
\label{compositepvalue}
\sim\ \ \left|\frac{|\mu|-|M_1|}{M_1}\right|,\ \ 
\left|\frac{|\mu|-|M_2|}{M_2}\right|
\ ,
\end{eqnarray}
respectively. They should be combined multiplicatively. It is easy to check from Fig.~\ref{fig:binowinohino_M1-M2p-Ann} that this leads to a fine-tuning which is typically an ${\cal O}(1-10)$ factor more severe than the tuning in the regions related to well tempered bino-Higgsino or bino-wino, as expected.

\section{Funnels}
\label{sec:funnels}
When the LSP can resonantly annihilate in the $s-$channel through an intermediate boson, say $F$, and $m_{\chi_1^0}\simeq m_F/2$, the annihilation cross-section increases enormously. In this way, scenarios of almost pure bino, which normally lead to excessive relic density, can be rescued. The funnel particle, $F$, can be the $Z$-boson, the ordinary Higgs boson, $h$, and the pseudoscalar, $A$. Note that for the first two cases $\chi_1^0$ must be rather light, which implies, as a matter of fact, that it should be nearly pure bino; otherwise, either $M_2$ or $\mu$ would be necessarily close to $m_Z/2$ or $m_h/2$, thus leading to charginos below the LEP limit, $M_{\chi^\pm}\gsim 100$ GeV. 
For the $A-$funnel case, it is desirable that $\chi_1^0$ be mostly bino as well, otherwise its mass should be very large. Notice here that, without the help of any funnel, the annihilation cross section of pure Higgsinos or pure winos is already quite efficient, which requires them to be rather heavy ($\simeq 1$ TeV and $\simeq 3$ TeV respectively) in order to reproduce the correct relic density. If, in addition, there is a channel of resonant annihilation (funnel), then their masses should be even larger, which would imply a heavier supersymmetric spectrum, and thus a more severe EW fine-tuning. 

Let us first consider the $A-$funnel, i.e. the resonant annihilation through the $A$ pseudoscalar
\be
\label{A-funnel}
\chi_1^0 \chi_1^0\rightarrow A\rightarrow SM\ {SM}, 
\ee
where $SM\ SM=b\bar b, gg$, etc. Note that, for this process to take place, $\chi_1^0$ must have a non-vanishing component of Higgsino, so that the $\chi_1^0-\chi_1^0-A$ vertex is in fact $\tilde B-\tilde H^0-A$. 
Consequently, the larger the Higgsino component of $\chi_1^0$ (and thereby the smaller $\mu$), the more efficient the annihilation. 
Another point to keep in mind is that, even if $M_1$ is below the resonant value, i.e. $m_A/2 - M_1 > \Gamma_A$, there can still be resonant annihilations, thanks to the thermal agitation in the early Universe, for some collisions the kinetic energy of the neutralinos can be large enough to reach $s\simeq m_A/2$. 
Of course, this amounts to a ``Boltzmann penalty" for the averaged cross section. On the other hand, if $M_1>m_A/2$, resonant annihilations are not possible. 
Then, the relic density, eq.~(\ref{Omegathermal}), as a function of $M_1$ shows a characteristic asymmetric dependence on $M_1$ in the resonance-region. 

\begin{figure}[ht]
  \centering
  $$\includegraphics[width=10cm]{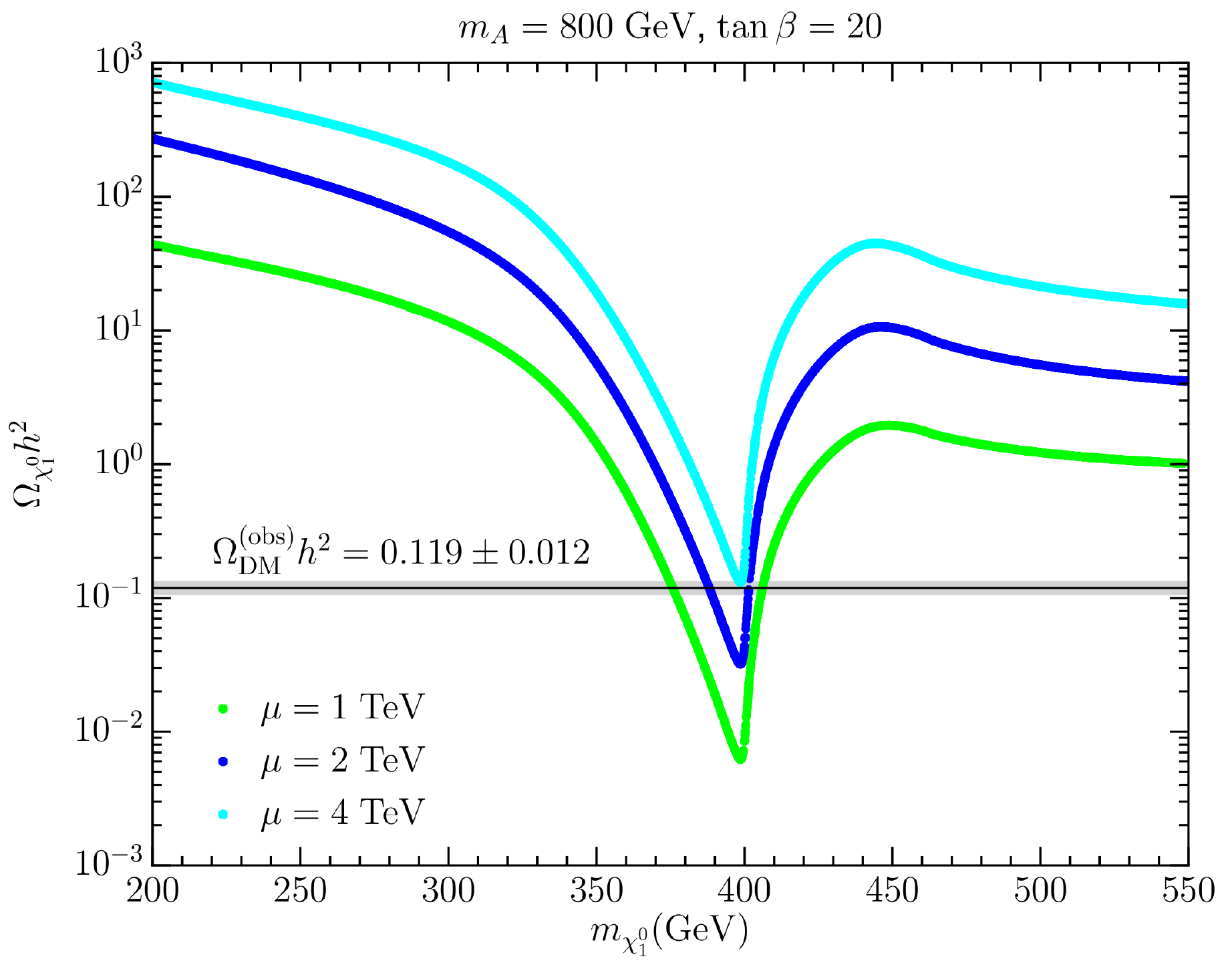} $$   
  \caption{$\Omega_{\chi_1^0} h^2$ vs $M_1$ in the $A-$funnel scenario for $m_A=800$ GeV, $\tan\beta=20$ and three (positive) values of $\mu$. The grey band denotes the observed relic abundance.}
  \label{fig:Afunnel_Mn1-Omega}
\end{figure}

All this is illustrated in Fig.~\ref{fig:Afunnel_Mn1-Omega} for $m_A=800$ GeV.
Note that ``far" from the resonant point the dependence of $\Omega_{\chi_1^0}$ on $M_1$ has an exponential shape due to the above-mentioned Boltzmann penalty. 
Thus we can expect that, similarly to what happened for the bino-wino co-annihilation scenario (section \ref{sec:binowino}), the standard criterion for the fine-tuning is not suitable here and typically overestimates the real fine-tuning. 
Nevertheless, when $M_1$ approaches the resonant point, we expect the opposite (recall the discussion around Fig.~\ref{fig:standard_criterion_1}). In the limiting case,
in which the physical region, $\Omega_{\chi_1^0}\simeq\Omega_{\rm DM}^{\rm (obs)}$, is close to the minimum of the curve
(this occurs for $\mu\simeq 4000$ GeV in the example of Fig.~\ref{fig:Afunnel_Mn1-Omega}), then the standard criterion indicates no fine-tuning at all; however this is obviously the most fine-tuned case!

In contrast, the $p-$value criterion is very transparent and easy to apply. Let us call $M_1^{\rm (0)}$ the value of $M_1$ that, for given $m_A$ and $\mu$, leads to $\Omega_{\chi_1^0}=\Omega_{\rm DM}^{\rm (obs)}$. Then, only in the narrow range $M_1 \in [M_1^{\rm (0)}, M_1^{\rm (0)}+\Delta m]$, with $\Delta m \simeq |m_A/2 - |M_1^{\rm (0)}||$,
the relic density will be equal or smaller than the observed value, as it is clear from Fig.~\ref{fig:Afunnel_Mn1-Omega}. Therefore, the $p-$value is (once more) simply $\Delta m/|M_1^{\rm (0)}|$.

\begin{figure}[ht]
  \centering
  $$\includegraphics[width=10cm]{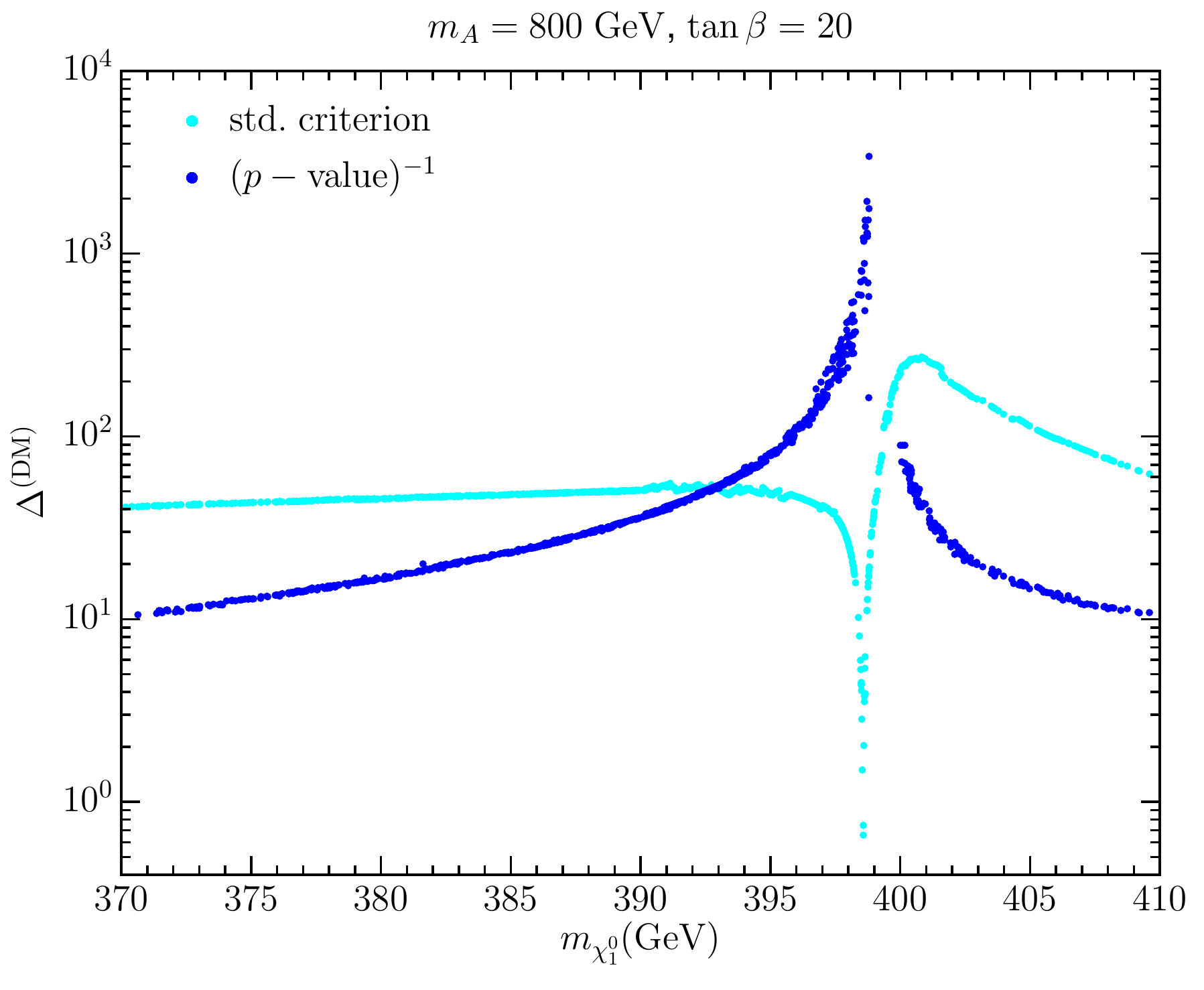} $$   
  \caption{Fine-tuning in the $A-$funnel scenario for $m_A=800$ GeV and $\tan\beta=20$, calculated using the ``standard criterion" [see eq.~(\ref{BGDM})] (cyan band) and using the $p-$value criterion, i.e. $\left(|m_A/2-|M_1||/|M_1|\right)^{-1}$ (blue band). For each value of $M_1$, the corresponding (positive) $\mu$ is chosen so that $\Omega_{\rm DM}^{\rm (obs)}$ is reproduced.}
  \label{fig:Afunnel_Mn1-FT}
\end{figure}

Fig.~\ref{fig:Afunnel_Mn1-FT} illustrates the previous discussion, showing the fine-tuning
calculated with the standard criterion, eq.~(\ref{BGDM}), and the one estimated as
the inverse of the $p-$value, eq.~(\ref{thetapvalue}), for $m_A=800$ GeV.  For each
value of $M_1$, the corresponding $\mu> 0$ is chosen so that the observed
relic density~(\ref{OmegaDM}) is fulfilled (recall that the value of  $\mu$
determines the amount of Higgsino mixing). 

\begin{figure}[ht]
  \centering
  $$\includegraphics[width=10cm]{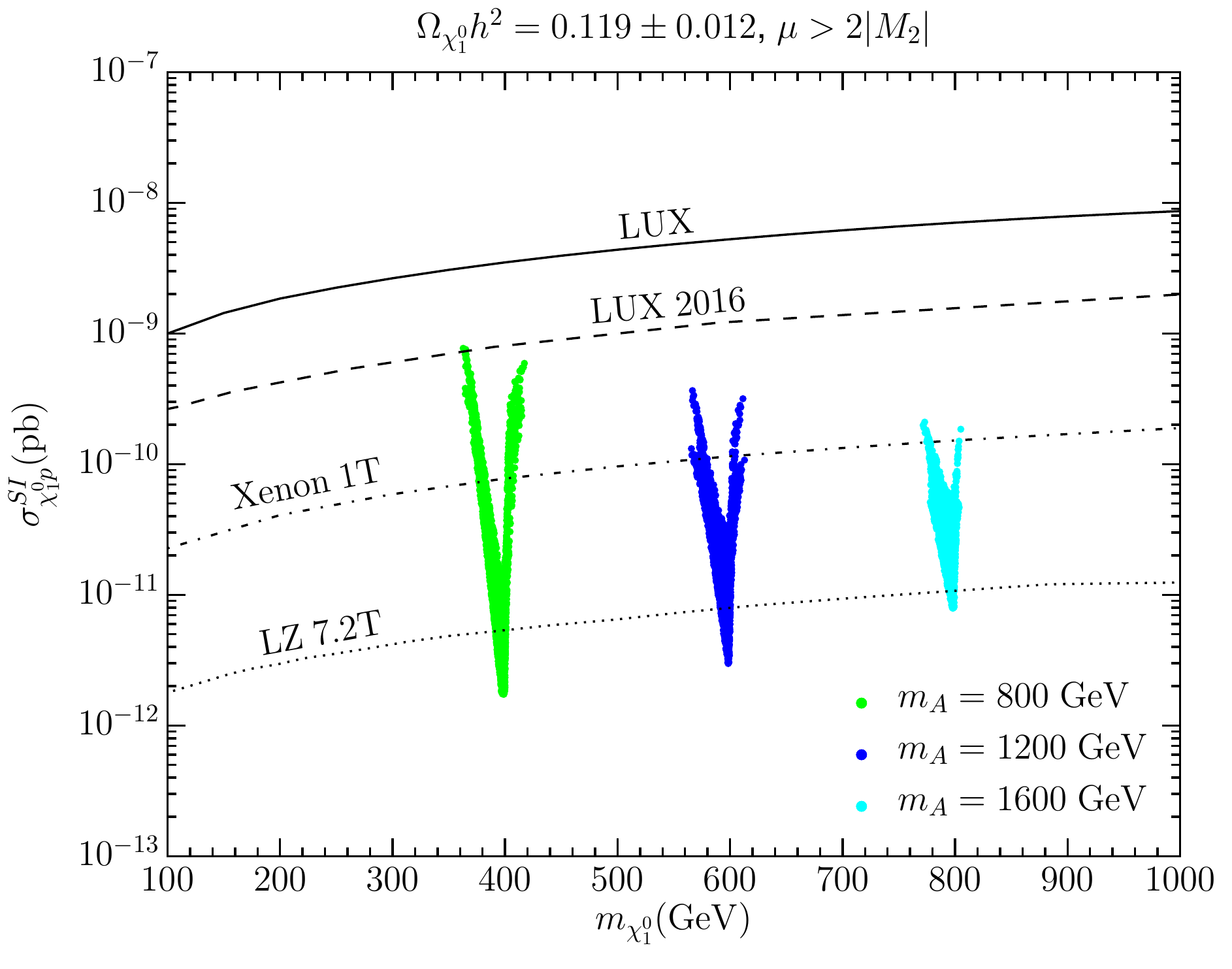} $$   
  \caption{Spin-independent neutralino-proton cross section in the $A-$funnel scenario for $m_A=800$ GeV. 
  The solid and dashed black lines denote the current LUX upper limit and the preliminary LUX 2016 bound, respectively. 
  The XENON 1T (dash-dotted line) and LZ (dotted line) projected sensitivities are also depicted.}
  \label{fig:Afunnel_Mn1-sSI}
\end{figure}

It is also worth-mentioning that this scenario is quite safe with respect to the current DM direct detection bounds because the elastic scattering cross section does not benefit from any resonant enhancement. This is shown in Fig.~\ref{fig:Afunnel_Mn1-sSI} for three different values of $m_A$.

Let us finally mention that for $|M_1| \lsim \frac{3}{4} (m_A/2)$ the enhancement due to the resonant annihilation is lost, but the relic density can still be reproduced if $\mu$ (and/or $M_2$) are close enough to $M_1$ for the scenario to become a well-tempered neutralino case. 
In that case, the fine-tuning is due to this well-tempered character and has been analyzed in sections \ref{sec:binoHiggs} and \ref{sec:binowino}.

\begin{figure}[ht]
  \centering
  $$\includegraphics[width=10cm]{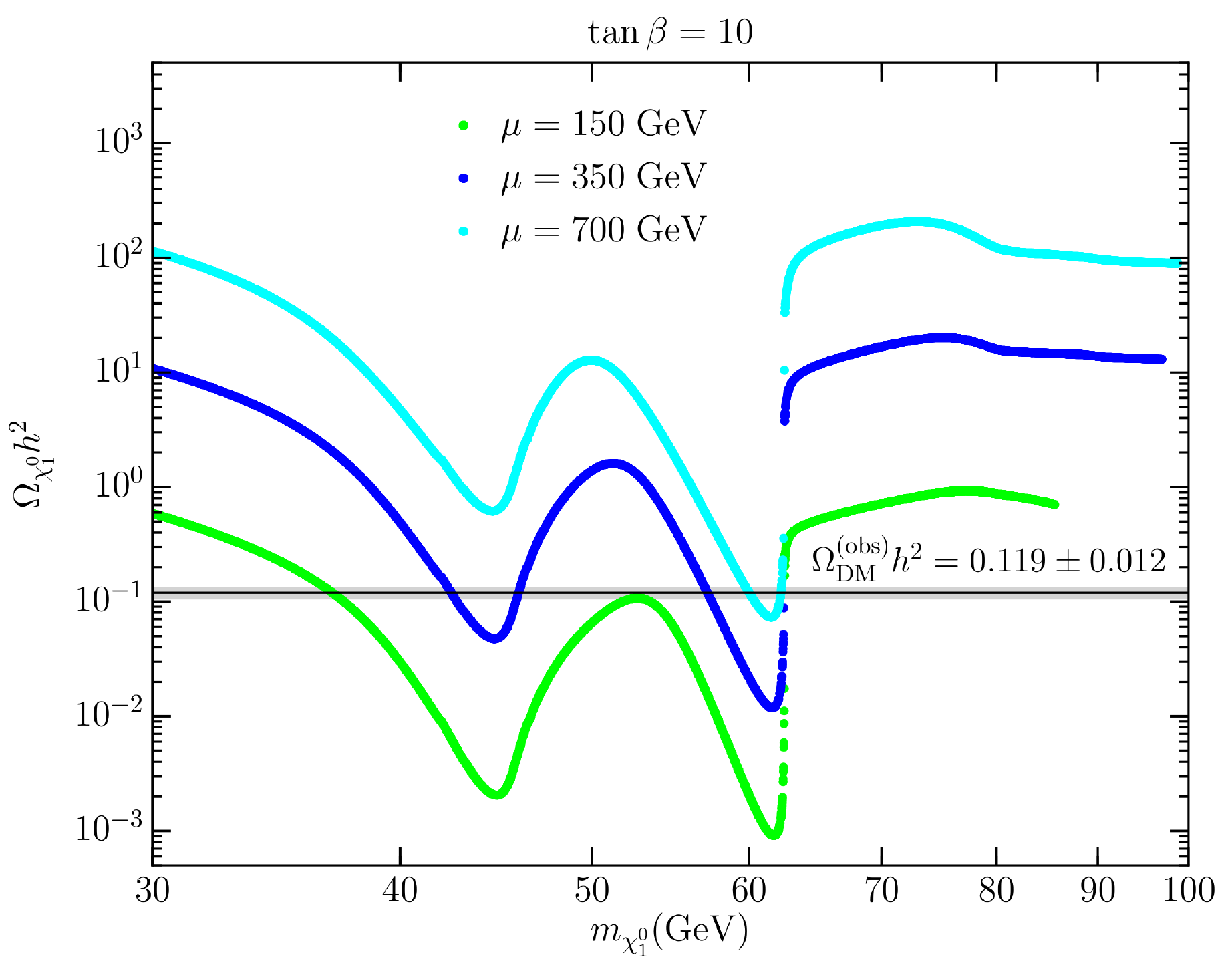} $$   
  \caption{$\Omega_{\chi_1^0} h^2$ vs $M_1$ in the region of $h-$ and $Z-$funnels for $\tan\beta=10$ and three (positive) values of $\mu$. The grey band denotes the observed relic abundance.}
  \label{fig:hfunnel_Mn1-Omega}
\end{figure}

Now, we turn to the $h-$ and $Z-$funnels
\bea
&\chi_1^0 \chi_1^0&\rightarrow h\rightarrow SM\ {SM}, 
\label{h-funnel}\\
&\chi_1^0 \chi_1^0&\rightarrow Z\rightarrow SM\ {SM}. 
\label{Z-funnel}
\eea
Similarly to the $A-$funnel case, these channels require $\chi_1^0$ to have a non-vanishing Higgsino component, so that the $\chi_1^0-\chi_1^0-h$ ($\chi_1^0-\chi_1^0-Z$) vertex has the $\tilde B-\tilde H^0-h$ ($\tilde H^0-\tilde H^0-Z$) 
structure\footnote{For the $h-$funnel $\sigma_{\rm ann} \propto |N_{11}N_{14}|^2$, while for the $Z-$funnel $\sigma_{\rm ann} \propto |N_{11}N_{13}- N_{11}N_{14}|^4$, with $N$ being the neutralino mass matrix.}. Thus, again, the larger the Higgsino component of $\chi_1^0$ (and thus the smaller $\mu$), the more efficient the annihilation. This effect is stronger for the $Z-$funnel, as it involves the Higgsino component in the two incoming neutralinos. All this is illustrated in Fig.~\ref{fig:hfunnel_Mn1-Omega}, which shows the dependence of $\Omega_{\chi_1^0}$ vs $M_1$ for three different values of $\mu$.

Regarding the fine-tuning issue, as for the $A-$funnel case, we expect that typically, the standard criterion overestimates the fine-tuning, except when the physical region, $\Omega_{\chi_1^0}\simeq\Omega_{\rm DM}^{\rm (obs)}$, is close  to a stationary point. 
Looking at Fig.~\ref{fig:hfunnel_Mn1-Omega}, there are now three stationary points, corresponding to the two minima at 
$m_{\chi_1^0}\simeq m_h/2, M_Z/2$ and to the maximum between both. For the very same reasons as for the $A-$channel, we find that 
a more robust and reliable measure of the fine-tuning is provided by the $p-$value $\simeq \Delta m /|M_1|$, where $\Delta m$ is the length of the $M_1-$range where the relic density is equal or smaller than the observed value\footnote{
Around the Higgs-resonance $\Delta m\simeq |m_h/2 - m_{\chi_1^0}|$, while around the $Z-$resonance $\Delta m\simeq 2|m_Z/2 - m_{\chi_1^0}|$, due to the larger width of the $Z-$boson. This can be appreciated in Fig.\ref{fig:hfunnel_Mn1-Omega}.
}.

\begin{figure}[ht]
  \centering
  \hspace{-0.8cm}
  \includegraphics[width=.5\linewidth]{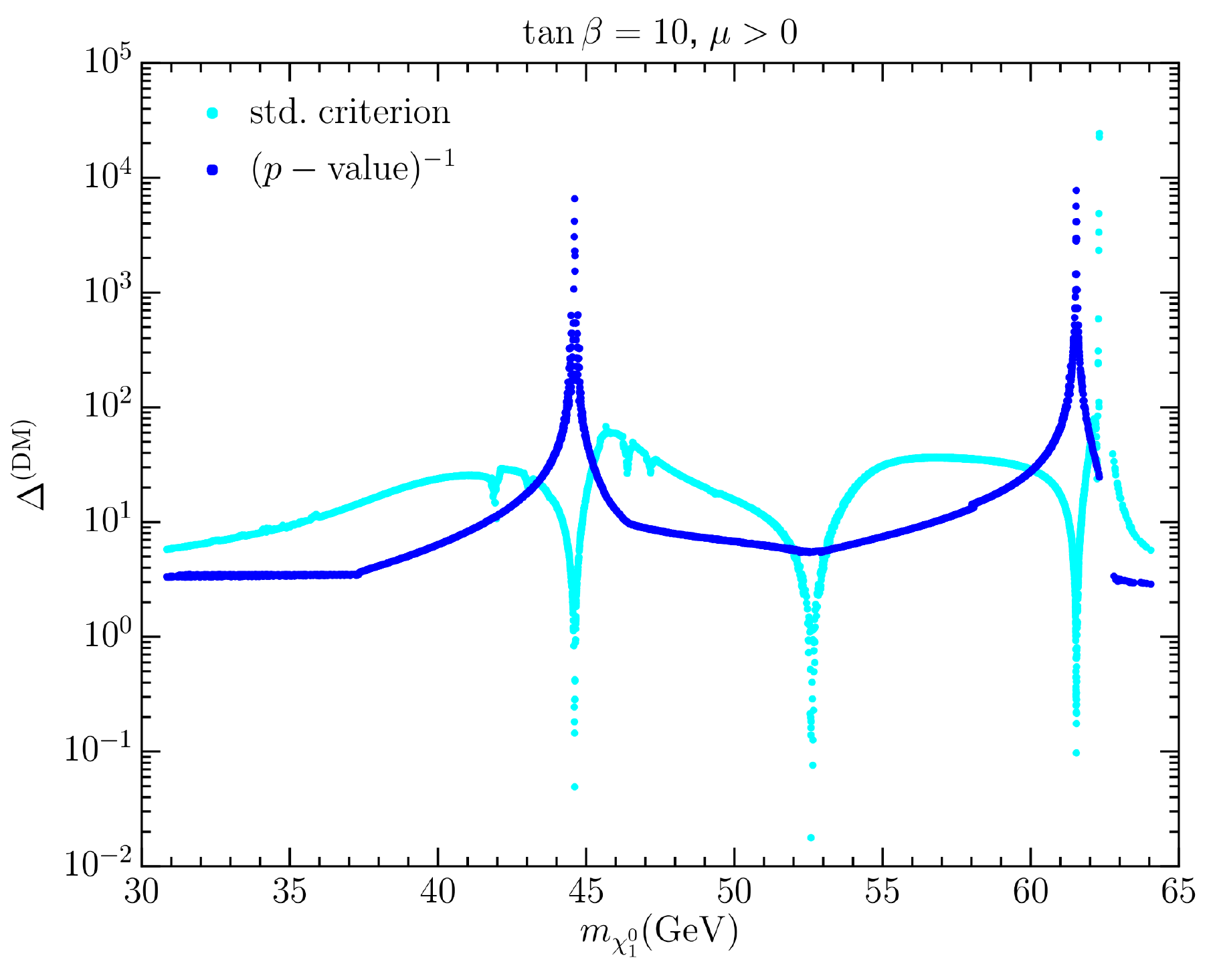}  
  \includegraphics[width=.5\linewidth]{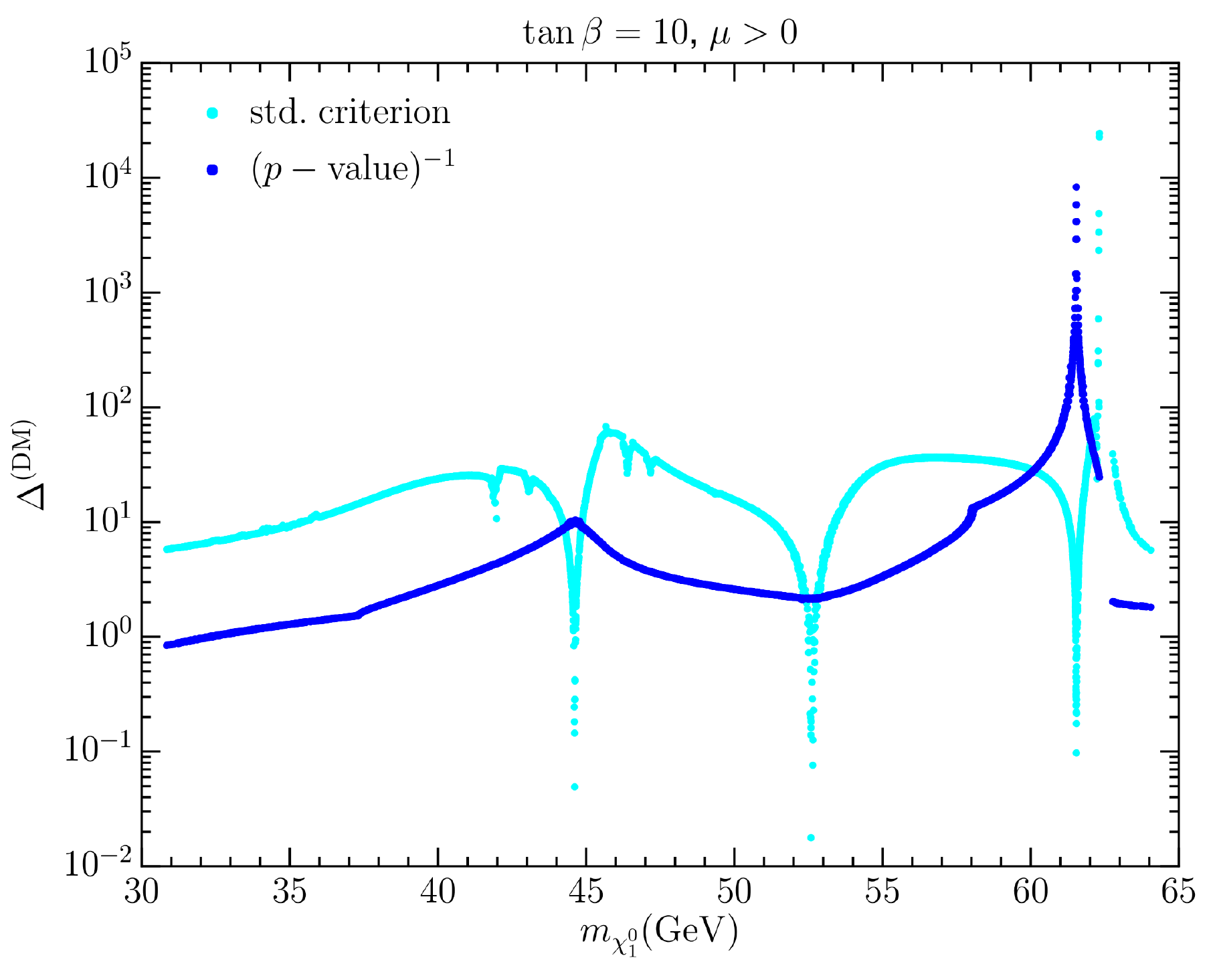}
  \caption{Left panel: Fine-tuning in the region of $h-$ and $Z-$funnels for $\tan\beta=10$, calculated using the ``standard criterion" [see eq.~(\ref{BGDM})] (cyan band) and using the $p-$value criterion, as explained in the text (blue band). Right panel: The same but adding up the $p-$values corresponding to the $h-$ and $Z-$funnels, when the value of $\mu$ allows for both possibilities (see text).}
  \label{fig:hfunnel_Mn1-FT}
\end{figure}

Fig.~\ref{fig:hfunnel_Mn1-FT} (left panel), which is similar to
Fig.~\ref{fig:Afunnel_Mn1-FT} but for the $h-$ and $Z-$funnels, illustrates
the previous discussion. Again, for each value of $M_1$, the corresponding
$\mu>0$ is chosen so that the observed relic density (\ref{OmegaDM}) is
reproduced.  As argued above, the standard criterion overestimates
  (non-dramatically) the fine-tuning in most of the $M_1-$range. However,
  around the three stationary points (in particular the two resonant points),
  it underestimates the fine-tuning dramatically. Indeed, apart from the resonant
points, the fine-tuning (estimated with the $p-$value criterion) is quite mild
($\lsim 10$). Note from Fig.~\ref{fig:hfunnel_Mn1-Omega} that for $150\ {\rm
    GeV}\lsim\mu\lsim 450\ {\rm GeV}$ both the $h-$funnel and the $Z-$funnel
 can successfully reproduce a relic density equal to the observed size, or even smaller, if $M_1$ is positive and has the appropriate value. Thus, the two $p-$values should be added,
implying a less severe fine-tuning.  The results are shown in
Fig.~\ref{fig:hfunnel_Mn1-FT} (right panel). Notice that in this way the peak
associated with the $Z-$funnel region is blown-up. This occurs because, for a
given value of (positive) $\mu$, the possibility of $Z-$funnel annihilation is
always accompanied by the possibility of $h-$funnel annihilation, but not the
other way round, see Fig.~\ref{fig:hfunnel_Mn1-Omega}. For $\mu<0$ the results
are similar, but in that case the blown-up peak corresponds to the $h-$funnel
for analogous reasons. The bottom line is that, apart from the peaks very
close to the resonant points, the $h-$ and $Z-$funnels show very mild or
non-significant fine-tuning.

 \section{Annihilation and Co-annihilation}
\label{sec:coann}
Co-annihilation occurs when one or several particles with masses close to the LSP annihilate efficiently. In that case, the relic density is still given by eq.~(\ref{Omegathermal}), but with the, effective, averaged annihilation cross-section, $\langle \sigma_{\rm eff}v\rangle$, given by eq.~(\ref{sigma_coan}). Once more, this mechanism is only useful if the LSP is essentially a bino, which is the instance where the LSP does not annihilate efficiently enough at early times.

Due to the Boltzmann factor in eq.~(\ref{sigma_coan}), $\langle \sigma_{\rm eff}v\rangle$, and thus $\Omega_{\chi_1^0}$, is exponentially sensitive to the mass gap, $\Delta m$, between the neutralino and its neighbouring particles. Hence, the most important dependence on $M_1$ goes, qualitatively, as
\begin{eqnarray}
\label{Omega_coan2}
\Omega_{\chi_1^0}\sim e^{-\xi x_f \left(\frac{\Delta m}{M_1}\right)}
\ ,
\end{eqnarray}
where $x_f\simeq 20$ and $\xi$ is typically ${\cal O}(1)$.
Some particularly important possibilities for the co-annihilating particles are the gluino, the stop and the stau (beside Higgsinos and winos, analyzed in previous sections). 
The above exponential dependence makes the standard criterion of fine-tuning to be quite severe in all cases, 
\begin{eqnarray}
\label{Delta_coan}
\Delta_{M_1} = \left|\frac{\partial \log \Omega_{\chi_1^0}}{\partial \log M_1}\right|\simeq \xi x_f\frac{\tilde m}{M_1} = {\cal O}(1) \times 20
\ ,
\end{eqnarray}
where $\tilde m$ is the mass of the co-annihilating particle. The puzzling, and suspicious, fact is that this estimation of the tuning does {\em not} depend on the mass difference $\tilde m-M_1$. It is essentially constant independently of how precisely $M_1$ should be close to $\tilde m$. Certainly, this is due to the fact that the standard criterion measures sensitivity rather than fine-tuning, and these are not always equivalent.

\begin{figure}[ht]
  \centering
  \includegraphics[width=.5\linewidth]{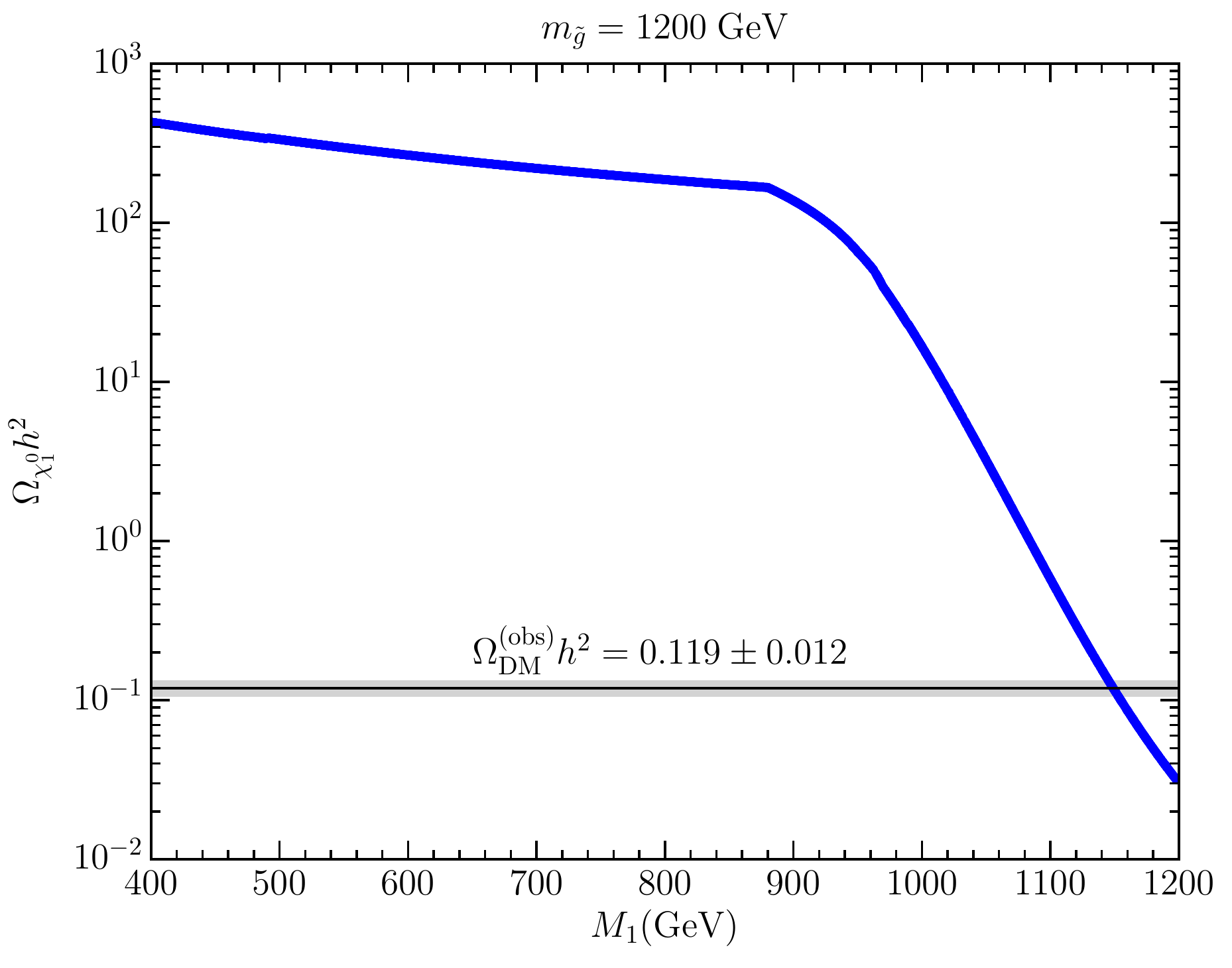}
  \includegraphics[width=.491\linewidth]{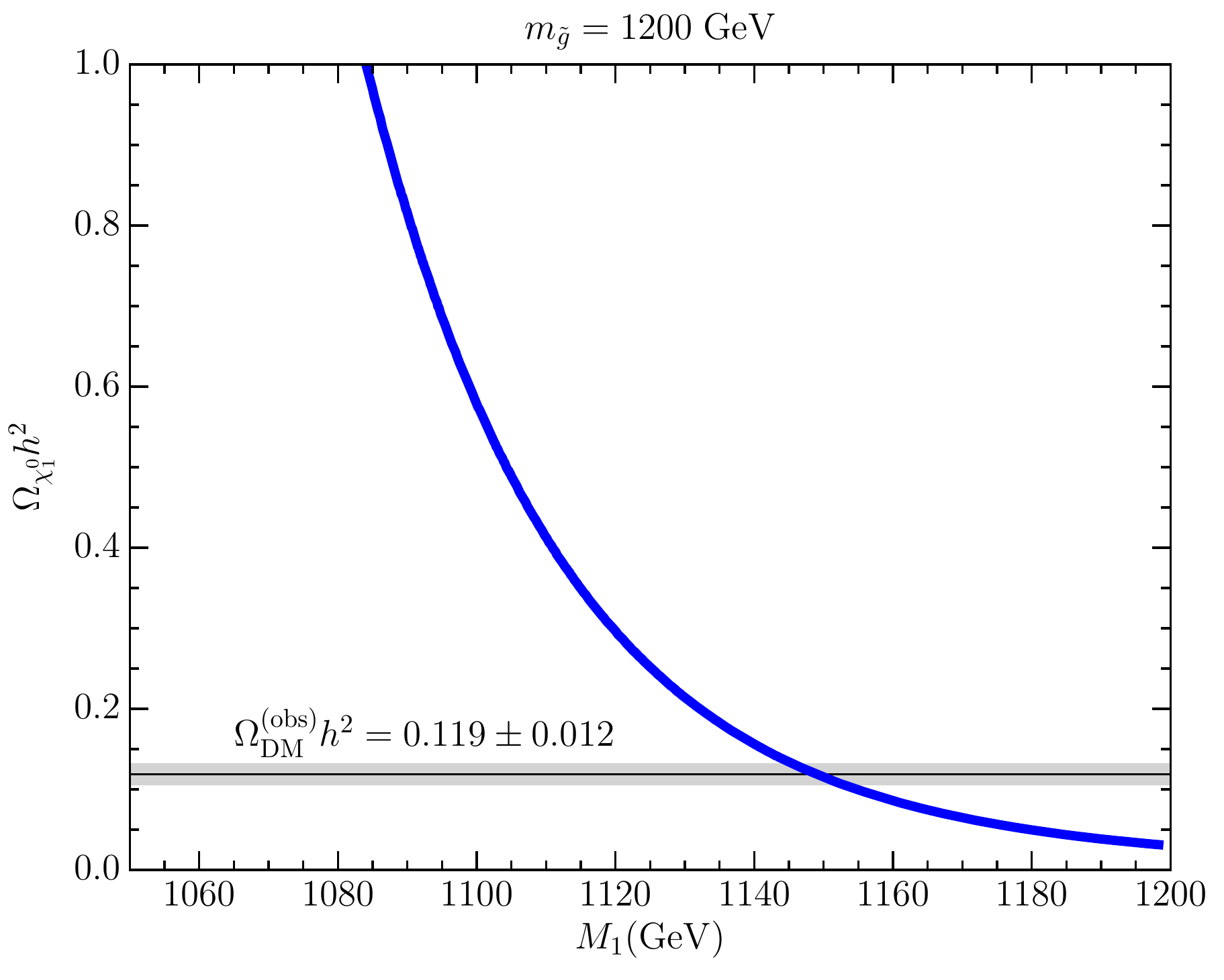}
  \caption{$\Omega_{\chi_1^0} h^2$ vs $M_1$ in the gluino co-annihilation scenario for fixed $m_{\tilde g}=1200$ GeV. The lightest neutralino is essentially bino. Left (right) panel shows the dependence in logarithmic (linear) units.}
  \label{fig:gluinoco_M1-Omega}
\end{figure} 

In order to illustrate these aspects, let us consider the case of gluino co-annihilation. Fig.~\ref{fig:gluinoco_M1-Omega} (left panel) shows $\Omega_{\chi_1^0}$ vs $M_1$ for $m_{\tilde g}=1200$ GeV.
The exponential dependence on $M_1$ in the co-annihilation region has been zoomed in linear scale in the right panel. Now, comparing this figure to Fig.~\ref{fig:standard_criterion_2}, it is clear that the standard criterion leads to an overestimation of the fine-tuning, since the truncation of $\Omega_{\chi_1^0}(M_1)$ at first order around the physical point is not good enough to describe the whole region where $\Omega_{\chi_1^0}\leq \Omega_{\rm DM}^{\rm (obs)}$.
Once again a more sensible measure is given by the $p-$value,
\begin{eqnarray}
\label{pvalue_coan}
p-{\rm value} \simeq \frac{\Delta m} {|M_1|} \ .
\end{eqnarray}
 
\begin{figure}[ht]
  \centering
  $$\includegraphics[width=10cm]{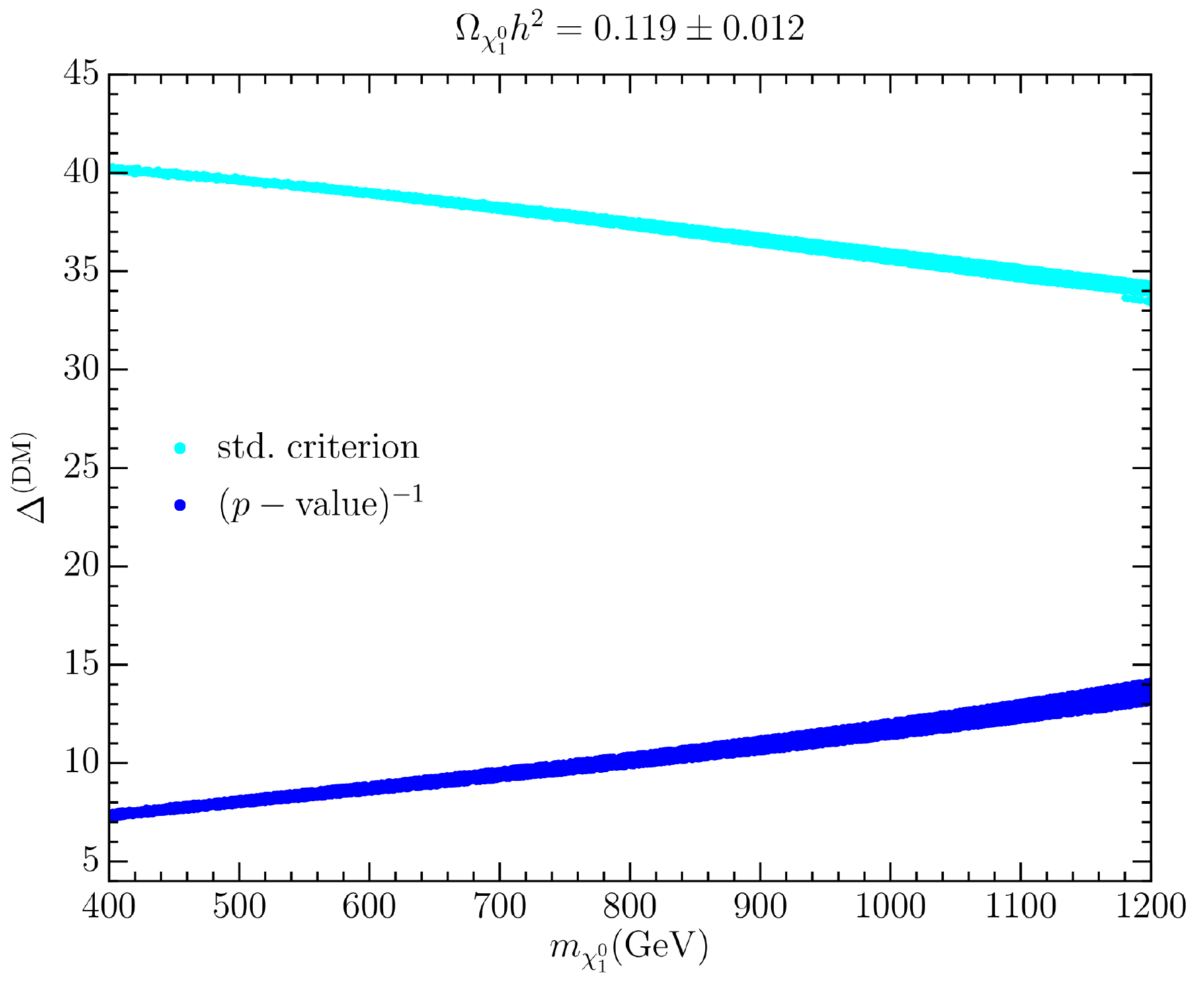} $$   
  \caption{Fine-tuning in the gluino co-annihilation scenario, calculated using the ``standard criterion" [see eq.~(\ref{BGDM})] (cyan band) and using the $p-$value criterion, i.e. $(\Delta m/M_1)^{-1}$ (blue band). The $m_{\tilde g}$ value is chosen so that the observed relic density, $\Omega_{\chi_1^0}h^2=0.119\pm 0.012$, is always reproduced. The lightest neutralino is essentially bino.}
  \label{fig:gluinoco_Mn1-FT}
\end{figure}

Fig.~\ref{fig:gluinoco_Mn1-FT} shows the fine-tuning calculated with the
standard criterion eq.~(\ref{BGDM}) and the one estimated as the inverse of the
$p-$value eq.~(\ref{pvalue_coan}) for gluino co-annihilation. For each value of
$M_1$, the corresponding $m_{\tilde g}$ is chosen so that the observed relic
abundance (\ref{OmegaDM}) is reproduced.
As expected, the standard criterion clearly overestimates the fine-tuning and
is suspiciously independent of $M_1$. On the contrary, the $p-$value criterion
shows a less severe tuning, especially for $M_1\lsim 500$ GeV, where it
becomes almost irrelevant. The increase in this fine-tuning with $M_1$
  occurs because the heavier the gluino, the less efficient becomes its
  annihilation, and this must be compensated by a more precise gluino-bino
  degeneracy.

Other co-annihilation cases, as the ones mentioned above, show a similar pattern.

\section{Connection to the electroweak fine-tuning}
\label{sec:EWFT}
The DM fine-tuning must be combined with the EW one, since both affect the same theoretical scenario, namely the MSSM. We recall here that the EW fine-tuning is reasonably well estimated by the ``standard measure", $\Delta^{\rm (EW)}={\rm max}\  \{{\partial m_h^2}/{\partial \theta_i}\}$ \footnote{This expression is equivalent to the expression (\ref{BG2}), once the radiative corrections to the Higgs effective potential are taken into account \cite{Casas:2014eca}.
}, where $\theta_i$ are the independent parameters of the model. 
As discussed in section \ref{sec:FT} [see eq.~(\ref{toy2})], this measure can be interpreted as the $p-$value associated with the small size of the EW scale. 
In sections \ref{sec:binoHiggs}--\ref{sec:coann}, we have evaluated the analogous $p-$value to reproduce $\Omega_{\rm DM}^{\rm (obs)}$. The main (computational) difference with the EW fine-tuning is that in this case the ``standard measure" of the fine-tuning is not a reliable estimation
of the $p-$value, so the latter has to be evaluated in a more direct way, as we have done. Typically, the EW fine-tuning is $\simgt {\cal O}(100)$, i.e. it is ${\cal O }(10)$ times more severe than the DM one, though the latter can be extremely larger at special places, see e.g. Figs.~\ref{fig:binohino_M1-FT},~\ref{fig:Afunnel_Mn1-FT} and~\ref{fig:hfunnel_Mn1-FT}. 
On the other hand, due to their common statistical interpretation (as $p-$values), it is clear that both fine-tunings should be multiplicatively combined.
A subtle aspect here is that the EW and DM fine-tunings arise from cancellations between the same set of parameters. 

This issue was analyzed in ref.~\cite{Casas:2005ev}, appendix A\footnote{The same prescriptions were independently found in later references \cite{Fichet:2012sn, Cheung:2012qy}.}. 
The idea is that when one computes the fine-tuning in a quantity, say $\Omega_{\rm DM}$, one is free to vary the input $\theta_i$'s only in a way that all the potential constraints (in this case the EW scale) are fulfilled. 
Denoting, for simplicity, $(\Delta_{\rm EW})_i$, $(\Delta_{\rm DM})_i$ the EW and DM fine-tunings (i.e. the inverse $p-$values) with respect to the $\theta_i$ parameters, and $G(\theta_i)=0$ the EW condition, one should project $\vec \Delta_{\rm DM}$ into the subspace orthogonal 
to the $G(\theta_i)=0$ hypersurface in the $\{\log \theta_i\}$ space. In other words, one has to re-define the DM fine-tuning as 
\be
\vec \Delta_{\rm DM}\rightarrow \vec \Delta_{\rm DM}-\frac{1}{|\vec \Delta G|^2}(\vec \Delta_{\rm DM}\cdot \vec \Delta G)\vec \Delta G\ ,
\ee
where $\overrightarrow {\Delta G}\equiv \{\partial G/\partial \log\theta_i\}\propto \vec \Delta_{\rm EW}$.

From this discussion, it is clear that only the parameters that contribute substantially to both the EW and DM fine-tunings are to play a relevant role in the previous projection.  In this sense, the EW fine-tuning is dominated by the initial values of $m_{\tilde t}^2, M_{3}$ and $\mu$ parameters (see e.g. ref.~\cite{Casas:2014eca}), while the DM fine-tuning is dominated by $M_1$, and, depending on the annihilation mechanism, by $\mu$, $M_2$, $m_A$ or $m_{\chi'}$, where $\chi'$ is a possible co-annihilating particle (gluino, stop, etc.). Consequently, only the DM fine-tuning associated with $\mu$ or $m_{\chi'}$ is subject to be lowered by the non-trivial ``interference" with the EW one. The conclusion is that the DM fine-tuning with respect to $M_1$ (which is the one computed in previous subsections) is always representative of the total DM fine-tuning and does not need to be corrected by the projection onto the subspace satisfying the EW condition.

As discussed in section \ref{sec:binoHiggs}, the DM fine-tuning is quite independent of the details of the MSSM scenario (whether it is CMSSM, NUHM, etc., or the value of the high-energy scale, $M_{HE}$). In this respect, it is a very robust feature of the MSSM. 
This fortunate circumstance does not occur for the EW fine-tuning: $\Delta^{\rm (EW)}$ is much more model-dependent, since it depends on the initial values of $M_{3}, m_{\tilde t}$, etc., and on the correlations between them (e.g. whether or not there is a universal scalar mass). It also depends on $M_{HE}$. 
Concerning this point, we can presume that $M_{HE}=M_X$, because a lower value for $M_{HE}$ would typically lead to a very light gravitino\footnote{The gravitino could be heavier than this naive expectation in theories with extra dimensions, where gravity is stronger, see e.g. ref.~\cite{Antoniadis:1992fh}.} ($m_{3/2}\sim (M_{\rm HE}/M_{\rm P}) m_0$), which would then play the role of the LSP, instead of the lightest neutralino, as assumed in this paper. 
In any case, it is clear that for every scenario for which the DM fine-tuning has been computed in sections \ref{sec:binoHiggs}--\ref{sec:coann}, there is not a unique value of $\Delta^{\rm (EW)}$; the latter depends on the details of the high-energy theory.  

Nevertheless, instead of $\Delta^{\rm (EW)}$ one can consider $\Delta^{\rm (EW)}_{\rm min}$,
i.e. the minimal EW fine-tuning.  Normally $\Delta^{\rm (EW)}$ is dominated by the
$M_3-$contribution or by the $\mu-$contribution\footnote{We are not
  considering here an (unknown) hypothetical scenario where all the soft terms
  and $\mu$ are theoretically correlated in such fortunate way that their
  contributions to $m_h^2$ nearly cancel, so that there is no
  fine-tuning!}. So one could just set $m_{\tilde g}$ at its
  experimental lower bound, $\sim 1.3$ TeV \cite{Aad:2014lra}, which amounts
  to $M_3\simeq 0.59$ TeV. This gives $\Delta^{\rm (EW)}_{M_3}={\cal O}(100)$ \cite{Casas:2014eca},
independently of the DM scenario, provided it can accommodate such light
gluino. However, this is not always the case, e.g., as mentioned in the introduction,
for pure-wino DM (which does not entail DM fine-tuning), $M_{\chi_1^0}\sim
m_{\tilde W}\simeq 3$ TeV. This necessarily implies a heavy gluino, $m_{\tilde
  g}>3$ TeV and, in turn, a much larger EW fine-tuning, near ${\cal
  O}(1000)$ (notice here that, parametrically, $\Delta^{\rm (EW)}_{M_3} \propto
M_3^2\propto m_{\tilde g}^2$). If the gaugino masses are unified at high
energy, the EW tuning is even larger, since $m_{\tilde g}\simeq 2.8 m_{\tilde
  W}$ \cite{Casas:2014eca}. Other DM scenarios that may demand a heavy gluino are co-annihilation
(with a particle different from gluino), $A-$funnel or well-tempered
bino-wino-Higgsino, whenever $M_{\chi_1^0}\gsim 1.3$ TeV. In
contrast, if the co-annihilation is with a gluino, this can be
  substantially lighter than 1.3 TeV, because it would be quite degenerate with
the LSP, and thus invisible at the LHC. Hence, the latter scenario would
reduce $\Delta^{\rm (EW)}_{M_3}$~!  Actually, from Fig.~\ref{fig:gluinoco_Mn1-FT},
we see that co-annihilation with a light gluino could improve both the DM and
the EW fine-tunings.

Regarding the $\mu-$contribution, this is $\Delta^{\rm (EW)}_{\mu}\simeq
(2\mu/m_h)^2$ (for further details see ref.~\cite{Casas:2014eca}), so for
large enough $\mu$, $\Delta^{\rm (EW)}$ becomes dominated by
$\Delta^{\rm (EW)}_{\mu}$. More precisely, this happens for $\mu\gsim
  (1/2.3)m_{\tilde g}$, in particular for $\mu\gsim 570$ GeV if $m_{\tilde g}$
  is close to its 1.3 TeV lower bound \cite{Aad:2014lra}. Consequently, the
above-mentioned DM scenarios that implied a large $\Delta^{\rm (EW)}_{M_3}$, imply
also an even larger $\Delta^{\rm (EW)}_{\mu}$. It is also worth mentioning that
for pure Higgsino DM (which does not amount to DM fine-tuning),
$M_{\chi_1^0}\sim\mu\simeq 1$ TeV, implying $\Delta^{\rm (EW)}_{\mu}\gsim{\cal
  O}(200)$.

From the previous discussion, it is clear that $\Delta^{\rm (DM)}$ should be kept as small as possible, preferably compatible with a non-fine-tuned situation, otherwise the combined fine-tuning will be above several thousands. This can be achieved in an obvious way if the DM is pure Higgsino or pure wino. Nonetheless, as mentioned, in the latter case the EW fine-tuning raises to $\gsim {\cal O}(1000)$ (for pure Higgsino it also grows but in a much milder way). Other cases that essentially imply no (or very mild) DM fine-tuning are: well-tempered bino-Higgsino (if $M_{\chi_1^0}$ is not around 500 GeV); Higgs, $Z$ and $A$ funnels when $M_{\chi_1^0}$ is not too close to (half) the resonance mass; and co-annihilation scenarios when 
$M_{\chi_1^0}$ is rather light, i.e. $\lsim$ 500 GeV.

Finally, in all the cases one has to ensure that {\em i)} the value of
$m_{H_u}^2$ at LE has the right size ($\sim \mu^2$) to enable the correct EW
breaking, and {\em ii)} the physical Higgs mass, $m_h\simeq 125$ GeV, is
reproduced. Both facts have to do with the values of $m_{\tilde t_L}^2$,
$m_{\tilde t_R}^2$, $A_t$, and $m_{H_u}^2$ at HE. In general, it will be
possible to arrange these parameters so that (in combination with
  $M_3\simeq 1.3$ TeV) they implement {\em i)} and {\em ii)} without
significantly affecting the value of $\Delta^{\rm (EW)}$. However, if there are
theoretical correlations between the initial soft terms, the {\em i)} and {\em
  ii)} conditions may imply further constraints on the theory and thereby an
increase in $\Delta^{\rm (EW)}$. Next, we illustrate this point by considering the
case of pure Higgsino DM (one of the preferred ones from the above discussion)
when the theory is some kind of constrained MSSM.

\section{Accommodating Higgsino DM in the MSSM}
\label{sec:HiggsinoDM}
If the LSP is close to a pure Higgsino (with mass $\simeq \mu\simeq 1$ TeV), the rest of the supersymmetric particles must be heavier, which imposes conditions on HE parameters. Assuming $M_{\rm HE}=M_{\rm X}$ in what follows, one gets \cite{Casas:2014eca}
\bea
&m_{\tilde g}&\simeq 2.22 M_3 > 1\  {\rm TeV}\ ,
\nonumber\\
&m_{\tilde W}&\simeq 0.8 M_2 > 1\  {\rm TeV}\ ,
\nonumber\\
&m_{\tilde B}&\simeq 0.43 M_1 > 1\  {\rm TeV}\ ,
\eea
where $M_i$ are the gaugino masses at the HE scale. If these are unified, $M_1=M_2=M_3\equiv M_{1/2}$, then from the last equation 
\bea
M_{1/2}\simgt 2.3\  {\rm TeV}\ ,
\eea
implying
\be
m_{\tilde g} > 5.16\  {\rm TeV}\ .
\ee
This large value of $M_{1/2}$ implies a huge EW fine-tuning, see eq.~(\ref{FTM1/2}) below.
Other supersymmetric masses are also forced to be very large, e.g. the average stop mass, $m_{\tilde t}^2\equiv \frac{1}{2}(m_{\tilde t_1}^2 +m_{\tilde t_2}^2)$, reads \cite{Casas:2014eca}
\begin{eqnarray}
\label{mtbound1}
\overline m_{\tilde t}^2 \simeq \frac{1}{2} (5.945M_3^2+0.679  m_{\tilde t_L}^2+0.611m_{\tilde t_R}^2 + 0.182M_2^2 - 0.307m_{H_u}^2 \cdots) +m_t^2 \ .
\end{eqnarray}
Therefore, assuming gaugino unification leads to $m_{\tilde t}\simgt 4$ TeV. Similarly, using the formulae from ref.~\cite{Casas:2014eca}, one gets $m_{\tilde L}\simgt 1.5$ TeV. The singlet sleptons are much less constrained, but they are forced anyhow to live above 1 TeV so that the Higgsino plays the LSP role. All the previous relations are much less restrictive if one gives up gaugino unification or considers lower values of $M_{\rm HE}$, though the latter possibility is disfavoured if the LSP is not the gravitino.

\vspace{0.2cm}
\noindent
One can consider now the EW minimization condition, which, at (moderately) large $\tan\beta$, reads
\be
\label{Min}
-\frac{m_h^2}{2}=\mu^2 +m_{H_u}^2 \ ,
\ee
with all  quantities defined at LE, in particular (see table 3 in appendix A of \cite{Casas:2014eca})
\bea
\label{mHu}
\left. m_{H_u}^2\right|_{LE}&=&-1.6M_3^2+0.63m_{H_u}^2-0.37m_{\tilde t_L}^2-0.29m_{\tilde t_R}^2
\nonumber\\
&&+0.28 A_t M_3 +0.2 M_2^2-0.13 M_2M_3-0.11 A_t^2+\cdots \ ,
\eea
where the variables in the r.h.s. are at HE. Notice from eq.~(\ref{Min}) that for $|\mu|\simeq 1$ TeV, $m_{H_u}^2 \simeq -|\mu|^2 \simeq -1\ {\rm TeV}^2$, which is an additional constraint. In fact, in the popular constrained-MSSM (CMSSM), it seems impossible to satisfy this constraint with all the supersymmetric masses higher than 1 TeV. 
Note that for the CMSSM, the contributions from $m_{H_u}^2$, $m_{\tilde t_L}^2$ and $m_{\tilde t_R}^2$ almost cancel in eq.~(\ref{mHu}), which is the well-known focus-point behaviour. 
Then, it is almost impossible to compensate the huge negative contribution coming from $M_3^2$ (recall that in the CMSSM $M_3=M_{1/2}\simeq 2.3$ TeV, due to gaugino unification). 
Using a very large $A_t$ with the appropriate sign does not help since the negative contribution from $A_t^2$ would dominate. 
Incidentally, pure wino DM is also unattainable, because whenever there is gaugino unification, the bino is lighter than the wino, so the latter cannot be the LSP.

Therefore, one has to go  beyond the CMSSM. The non-universal-Higgs-Mass model (NUHM) is like the CMSSM, but allowing the soft Higgs masses, $m_{H_u}^2, m_{H_d}^2$ to be different from the other scalar masses at HE (a usual choice is  $m_{H_u}^2=m_{H_d}^2$). 
Then, if $m_{H_u}^2$ is large enough at HE, one can achieve $m_{H_u}^2(LE) \simeq -1\ {\rm TeV^2}$ in eq.~(\ref{mHu}). This implies that the extra Higgs states are quite heavy (if the $m_{H_u}^2=m_{H_d}^2$ condition is imposed). The whole spectrum seems beyond the LHC. 
In addition, the model presents a high EW fine-tuning:
\be
\label{FTM1/2}
\Delta^{\rm (EW)}_{M_{1/2}}=\left|\frac{d \log m_h^2}{d\log M_{1/2} }\right| 
\simeq \left|-4\frac{M_{1/2}^2}{m_h^2}\left(-1.6+0.2-0.13 +0.14 \frac{A_t}{M_{1/2}}+\cdots\right)\right|\simeq 2000 \ .
\ee

Another possibility is to start with non-universal gaugino masses. This is a much more flexible scenario and, in principle, it does not seem difficult in this case to achieve the LSP condition for the Higgsino and the correct EW breaking with supersymmetric masses (in particular gluino masses) not far from their experimental lower bounds. 

\section{Conclusions}
\label{sec:conclu}

One of the most celebrated bonuses of supersymmetric theories is the presence of stable WIMPs, which are natural candidates for dark matter (DM). In the MSSM, such role is usually played by the lightest supersymmetric particle (LSP), which is typically the lightest neutralino. However, when one goes into the details, it turns out that in most scenarios some kind of tuning is needed in order to get $\Omega_{\rm DM}$ of the right magnitude. This fine-tuning is worrisome since it has to be combined with the ubiquitous electroweak (EW) fine-tuning problem, i.e. the delicate balance between soft terms required to reproduce the smallness of the EW scale.

Taking into account that the original motivation for low-energy SUSY was to solve the hierarchy problem, which is the EW fine-tuning problem of the SM, it is logical to demand SUSY scenarios to be as natural as possible. In this sense, there exists a vast literature examining the EW fine-tuning problem, but little concerning the DM one. 

In this paper, we study this problem in an, as much as possible, exhaustive and rigorous way. We have considered the MSSM framework, assuming that the LSP is the lightest neutralino, $\chi_1^0$, and explored various possible scenarios. 
These include different masses and compositions of $\chi_1^0$, which are completely defined by the parameters involved in the neutralino mass matrix ($M_1,M_2,\mu,\tan\beta$), 
as well as different mechanisms for neutralino annihilation in the early Universe (well-tempered neutralinos, funnels and co-annihilation scenarios). 
We also present a discussion about the statistical meaning of the fine-tuning and how it should be computed for the DM relic abundance, and combined with the EW fine-tuning. It turns out that the ``standard measurement" of fine-tuning, $\Delta = {d \log \Omega_{\rm DM}}/{d \log \theta}$ is not appropriate in most of the cases, and one has to evaluate the $p-$value associated with the smallness of $\Omega_{\rm DM}$, which, actually, amounts normally to a simpler computation.
A fortunate fact is that  the relevant (low-energy) parameters, involved in the neutralino mass matrix are essentially in one-to-one multiplicative correspondence with the initial (high-energy) parameters. 
This allows to compute the fine-tuning directly on the low-energy parameters with full generality. In consequence, the DM fine-tuning is quite independent of the details of the MSSM scenario (whether it is CMSSM, NUHM, etc., or the value of the high-energy scale). In this sense, it is a very robust feature of the MSSM. In contrast, 
the EW fine-tuning is much more model-dependent. 

Concerning the results, the fine-tuning related just to the DM relic abundance is negligible or very mild in a number of scenarios. More precisely,
when $\chi_1^0$ is essentially a pure Higgsino or a pure wino there is no fine-tuning associated with the DM relic density. 
Other cases that essentially imply no (or very mild) DM fine-tuning are: well-tempered bino-Higgsino (if $M_{\chi_1^0}$ is not around 500 GeV); Higgs, $Z$ and $A$ funnels when $M_{\chi_1^0}$ is not too close to (half) the resonance mass; and co-annihilation scenarios when 
$M_{\chi_1^0}$ is rather light, i.e. $\lsim$ 500 GeV.

Nevertheless, this is not the end of the story, as the DM fine-tuning must be
combined with the EW one. Modulo some subtleties discussed in this paper, both
fine-tunings should be essentially multiplicatively combined. Thus, one should
demand $\Delta^{\rm (EW)}$ to be as mild as possible. Normally $\Delta^{\rm (EW)}$ is
dominated by the $M_3-$contribution or by the $\mu-$contribution. So one could
just set $m_{\tilde g}$ at its experimental lower bound, $\sim 1.3$
  TeV, which leads to $\Delta^{\rm (EW)}={\cal O}(100)$, independently of the DM
scenario, provided it can accommodate such light gluino \cite{Casas:2014eca}. However, this
is not always the case. E.g. for pure-wino DM (which does not entail DM
fine-tuning), $M_{\chi_1^0}\sim m_{\tilde W}\simeq 3$ TeV. This necessarily
implies a heavier gluino and, in turn, a much larger EW fine-tuning, near
${\cal O}(1000)$. By contrast, if the co-annihilation is with a gluino, the
latter can be substantially lighter than 1.3 TeV, since it would be
invisible at the LHC. Hence, the latter scenario would reduce $\Delta^{\rm (EW)}$
!

As a final remark, naturalness is a reasonable guide to look for plausible supersymmetric scenarios. In this regard, a strong emphasis has been put on the EW fine-tuning, but the DM fine-tuning is also very important, as shown in this paper, especially when it is combined with the EW one. This feature should be taken into account when one explores ``natural SUSY" scenarios and their possible signatures at the LHC and in DM detection experiments. 

\newpage

\section*{Acknowledgements}
This work has been partially supported by the MICINN, Spain, under contract
FPA2013-44773-P, Consolider-Ingenio CPAN CSD2007-00042, as well as MULTIDARK
CSD2009-00064.  We also thank the Spanish MINECO {\em Centro de excelencia
  Severo Ochoa Program} under grant SEV-2012-0249.  The work of A.D. is partly
supported by the National Science Foundation under grant PHY-1520966.  S.R.
is supported by the Campus of Excellence UAM+CSIC. The work of M.E.C. was
supported by Funda\c{c}\~ao de Amparo \`a Pesquisa do Estado de S\~ao
Paulo.  R. RdA is supported by the Ram\'on y Cajal program of the Spanish MICINN and also thanks the support by the ``SOM Sabor y origen de la Materia" (FPA2011-29678), the ``Fenomenologia y Cosmologia de la Fisica mas alla del Modelo Estandar e lmplicaciones Experimentales en la era del LHC" (FPA2010-17747) MEC projects, the Severo Ochoa MINECO project SEV-2014-0398,  
the Consolider-Ingenio 2010 programme under grant MULTIDARK CSD2009-00064 and by the Invisibles European ITN project FP7-PEOPLE-2011-ITN, PITN-GA-2011-289442-INVISIBLES.

\bibliography{CCDRR}

\end{document}